\documentclass[]{aa}  
\usepackage[utf8]{inputenx}

\usepackage{graphicx}
\usepackage{fixltx2e}
\usepackage{url}

\usepackage{bm}
\usepackage{tikz}
\usetikzlibrary{patterns}
\usepackage{grffile}

\usepackage{txfonts}
\bibpunct{(}{)}{;}{a}{}{,} 
\begin{document} 

\title{LSDCat: Detection and cataloguing of emission-line sources in
  integral-field spectroscopy datacubes\thanks{ The \texttt{LSDCat}
    software is available for download at
    \texttt{http://muse-vlt.eu/science/tools} and via the Astrophysics
    Source Code Library at \texttt{http://ascl.net/1612.002}.}}

\titlerunning{LSDCat}
\authorrunning{E.~C.~Herenz \& L.~Wisotzki}

\author{Edmund~Christian~Herenz\inst{\ref{inst1},}\inst{\ref{inst2}}\and
        Lutz~Wisotzki\inst{\ref{inst1}}}

\institute{Leibniz-Institut für Astrophysik Potsdam (AIP), An der
  Sternware 16, 14482 Potsdam, Germany \label{inst1}
  \and
Department of Astronomy, Stockholm University, AlbaNova University Centre, SE-106 91,
Stockholm, Sweden \\ \email{christian.herenz@astro.su.se} \label{inst2}
}

   \abstract{We present a robust, efficient, and user-friendly
     algorithm for detecting faint emission-line sources in large
     integral-field spectroscopic datacubes  together with the public
     release of the software package \texttt{LSDCat} (\textit{Line
       Source Detection and Cataloguing}).  \texttt{LSDCat} uses a
     three-dimensional matched filter approach, combined with thresholding
     in signal-to-noise, to build a catalogue of individual line
     detections. In a second pass, the detected lines are grouped into
     distinct objects, and positions, spatial extents, and fluxes of
     the detected lines are determined. \texttt{LSDCat} requires only
     a small number of input parameters, and we provide guidelines for
     choosing appropriate values. The software is coded in Python and
     capable of processing very large datacubes in a short time.  We
     verify the implementation with a source insertion and recovery
     experiment utilising a real datacube taken with the MUSE
     instrument at the ESO \emph{Very Large Telescope}.  }

\keywords{methods: data analysis -- techniques: imaging spectroscopy}

\maketitle

\section{Introduction}
\label{sec:lsdc_intr}

One motivating driver for the construction of the current generation
of optical wide-area integral-field spectrographs such as the Multi
Unit Spectroscopic Explorer at the ESO VLT \citep[MUSE, in
operation since 2014;][]{Bacon2014,Kelz2015} or the Keck Cosmic Web Imager
\citep[KCWI, under construction;][]{Martin2010} is the detection of
faint emission lines from high-redshift galaxies. The high-level
data products from those instruments are three-dimensional (3D) 
arrays containing intensity-related values with two spatial axes and 
one wavelength axis (usually referred to as datacubes). So far, efficient, robust, and user-friendly detection and cataloguing software 
for faint-line-emitting sources in such datacubes is not publicly available.  
To remedy this situation we now present the \emph{Line Source Detection 
and Cataloguing} (\texttt{LSDCat}) tool, developed in the course of 
our work within the MUSE consortium.

Automatic source detection in two-dimensional (2D) imaging data
is a well-studied problem.  Various methods to tackle it have been
implemented in software packages that are widely adopted in the
astronomical community (see reviews by \citealt{Bertin2001} and
\citealt{Masias2012}, or the comparison of two frequently used tools
by \citealt{Annunziatella2013}).  A conceptually simple approach
consists of two steps: First, the observed imaging data is
transformed in order to highlight objects while simultaneously
reducing the background noise.  A particular transformation that
satisfies both requirements is the matched filter (MF) transform. Here
the image is cross-correlated with a 2D template that matches the 
expected light distribution of the sources to be detected.  Mathematically, 
it can be proven that for stationary noise the MF maximises the
signal-to-noise ratio ($S/N$) of a source that is optimally
represented by the template
\citep[e.g.][]{Schwartz1975,Das1991,Zackay2015,Vio2016}.  In the
second step, the MF-transformed image is segmented into objects via
thresholding, that is,\ each pixel in the threshold mask is set to 1 if 
the MF-transformed value is above the
threshold, and 0 otherwise. Connected 1-valued pixels then define the
objects on which further measurements (e.g. centroid coordinates,
brightnesses, ellipticities etc.)  can be performed.  Other image
transformations (e.g. multi-scale methods) and detection strategies
(e.g. histogram-based methods) exist, but the 
``MF + thresholding''-approach is most frequently employed, especially 
in optical/near-infrared imaging surveys for faint extragalactic
objects.  This widespread preference is certainly attributable to the
conceptual simplicity and robustness of the method, despite known limitations
\citep[see e.g.][]{Akhlaghi2015}.  But it is
also due to the availability of a stable and user-friendly
implementation of a software based on
this approach \citep{Shore2009}: \texttt{SExtractor}
\citep{Bertin1996}.

The detection and cataloguing of astronomical sources in 3D datasets
has so far mostly been of interest in the domain of radio astronomy.
Similar to integral-field spectroscopy (IFS), these observations result in 3D
datacubes containing intensity values with two spatial axes and one
frequency axis.  Here, especially surveys for extragalactic 21\,cm
\ion{H}{i} emission are faced with the challenge to discover faint
line emission-only sources in such datacubes.  The current generation of
such surveys utilises a variety of custom-made software for this task, 
also relying heavily on manual inspection of the datacubes.  Notably, the
approach of \cite{Saintonge2007} tries to minimise such error-prone
interactivity by employing a search technique based on matched
filtering, although only in spectral direction.  More recently, driven
mainly by the huge data volumes expected from future generations of
large radio surveys with the Square Kilometre Array, development and
testing of new 3D source-finding techniques has started
\citep[e.g.][]{Koribalski2012,Popping2012,Serra2012,Jurek2012}.
Currently, two software packages implementing some of these techniques
are available to the community: \textsc{duchamp} \citep{Whiting2012}
and \textsc{SoFiA} \citep{Serra2015}.  While in principle these
programs could be adopted for source detection purposes in optical
integral-field spectroscopic datasets, in practice there are
limitations that necessitate the development of a dedicated IFS
source detector. For example, the noise properties of long exposure
IFS datacubes are dominated by telluric line- and continuum emission
and are therefore varying with wavelength, in contrast to the more 
uniform noise properties of radio datacubes.  Moreover, the search
strategies implemented in the radio 3D source finders are tuned to
capture the large variety of signals expected.  But, as we argue in 
this paper, the emission line signature from compact high-redshift 
sources in IFS data is well described by a simple template that,  
however, needs an IFS-specific parameterisation. 
Finally, the input data as well as the parameterisation of detected sources 
is different in the radio domain compared to the requirements in 
optical IFS: typically radio datacubes have their spectral axis expressed 
as frequency and flux densities measured in Jansky, while in IFS cubes 
the spectral axis is in wavelengths and flux densities are measured as 
$f_\lambda$ with units of erg\,s$^{-1}$\,cm$^{-2}$\,\AA{}$^{-1}$.

\texttt{LSDCat} is part of a long-term effort, initially motivated by
the construction of MUSE, to develop source detection algorithms for
wide-field IFS datasets \citep[e.g.][]{Bourguignon201232}.  As part of
this effort, \cite{Meillier2016} recently released the \emph{Source
Emission Line FInder} SELFI.  This software is based on a Baysian
scheme utilising a reversible jump Monte Carlo Markov Chain
algorithm.  While this mathematically sophisticated machinery is
quite powerful in unearthing faint emission line objects in a datacube, 
the execution time of the software is too long for practical use.  
In contrast, the algorithm of \texttt{LSDCat} is relatively simple and 
correspondingly fast in execution time on state-of-the art workstations. 
It is also robust, as it is based on the matched-filtering method that has
long been successfully utilised for detecting sources in imaging data.
\texttt{LSDCat} is therefore well suited for surveys exploiting even 
large numbers of wide-field IFS datacubes.
Furthermore, the short execution time also permits extensive 
fake source insertion experiments to empirically reconstruct the
selection function of such surveys.

This article is structured as follows: In
Sect.~\ref{sec:method-description}, we describe the mathematical basis
and the algorithm implemented in \texttt{LSDCat}.  In
Sect.~\ref{sec:validation-software}, we validate the correctness of the
implementation in our software.  We then provide guidelines for the
\texttt{LSDCat} user for adjusting the free parameters governing the
detection procedure in Sect.~\ref{sec:gudel-optim-param} and conclude
in Sect.~\ref{sec:conclusion} with a brief outlook on future
improvements of \texttt{LSDCat}.  A link to the software repository is
available from the Astrophysics Source Code Libary
\url{http://ascl.net/1612.002}.  In Appendix \ref{sec:usage-example},
we provide a short example on how the user interacts with
\texttt{LSDCat} routines. The public release of the software includes
a detailed manual, to which we refer all potential users for details
of installing and working with the software.

\section{Method description \& implementation}
\label{sec:method-description}

\begin{figure}
  \centering
  \resizebox{\hsize}{!}{\includegraphics[trim=1cm 3cm 5cm 1cm, clip]{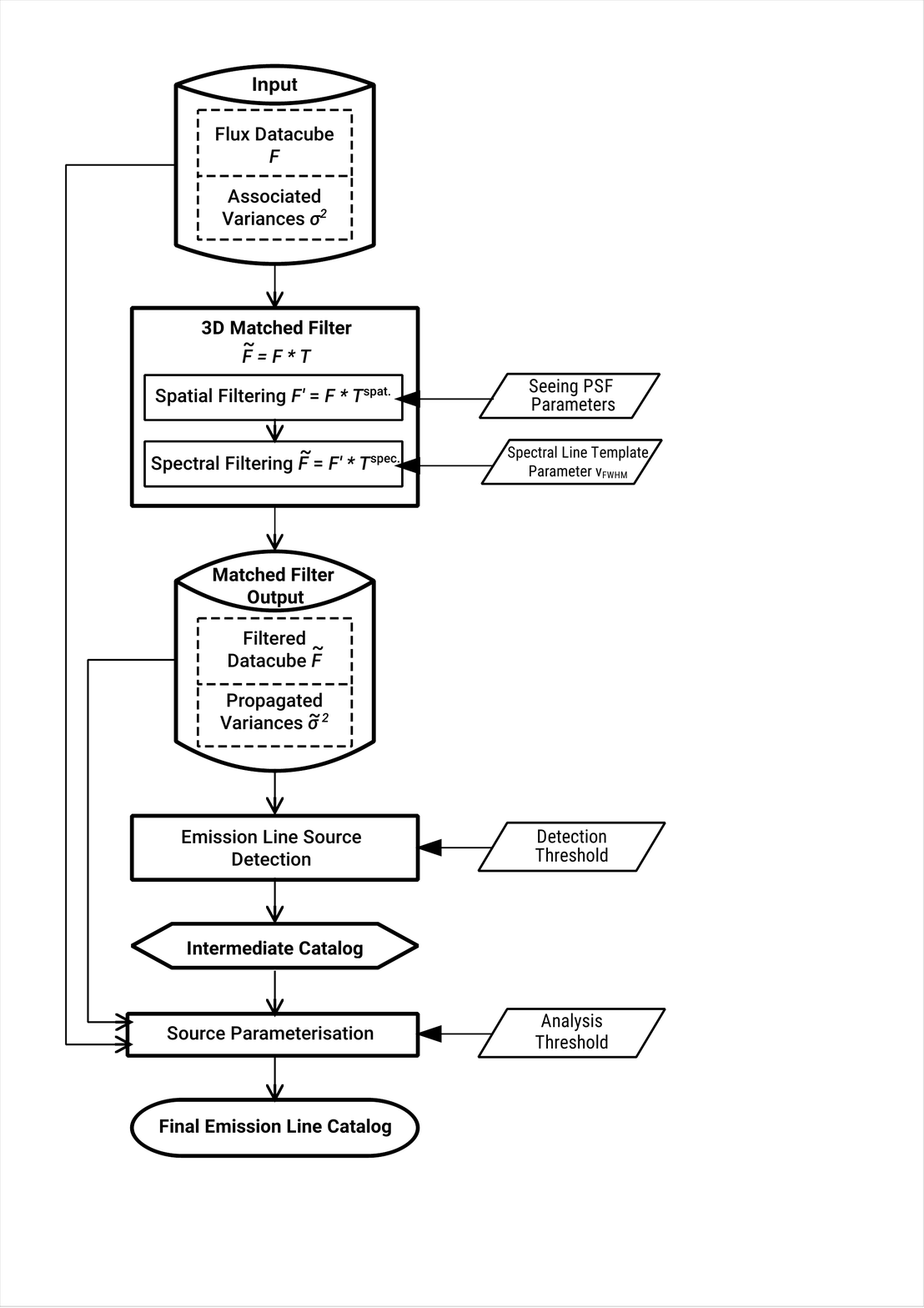}}
  \caption{Flowchart illustrating the processing steps of
    \texttt{LSDCat} from an input datacube to a catalogue of
    positions, shape parameters, and fluxes of emission line sources.
  }
  \label{fig:lsdcat_flow}
\end{figure}

\begin{table*}
  \caption{Routines of \texttt{LSDCat} with the required input
    parameters.}
  \centering
  \begin{tabular}{lll} \hline\noalign{\smallskip}
    Processing step & \texttt{LSDCat} Routine & Input parameters \\
    \noalign{\smallskip}\hline\noalign{\smallskip}
    Spatial filtering & \texttt{lsd\_cc\_spatial.py} & --\; PSF functional form: Moffat or Gaussian \\ 
    {}                & {}                           & --\; $p_0$,$p_1$,$p_2$ and $\lambda_0$ for \\
    {}                & {}                           & \phantom{--\;} $\mathrm{FWHM}(\lambda)\,\mathrm{[\arcsec{}]}=p_0 + p_1 (\lambda - \lambda_0) + p_2  (\lambda - \lambda_0)^2$ \\
    {}                & {}                           & --\; (Moffat $\beta$, if Moffat PSF) \\
    Spectral filtering& \texttt{lsd\_cc\_spectral.py}& --\; FWHM of Gaussian profile: $v_\mathrm{FWHM}$\,[km\,s$^{-1}$] \\
    Thresholding      & \texttt{lsd\_cat.py}         & --\; Detection threshold: S/N$_\mathrm{det.}$ \\
    Measurements      & \texttt{lsd\_cat\_measure.py} &--\; Analysis threshold: S/N$_\mathrm{ana.}$ \\
\noalign{\smallskip}\hline
  \end{tabular}
  \label{tab:lsdcat_routines}
\end{table*}

\subsection{Input data}
\label{sec:input} 

The principal data product of an integral field unit (IFU) observing
campaign is a datacube $\bm{F}$
\citep[e.g.][]{Allington-Smith2006,Turner2010}.  The purpose of
\texttt{LSDCat} is to detect and characterise emission-line sources in
$\bm{F}$.  We adopt the following notations and conventions: $\bm{F}$
is a set of volume pixels (voxels) $F_{x,y,z}$ with intensity related
values, for example, flux densities in units of
erg\,s$^{-1}$\,cm$^{-2}$\,\AA{}$^{-1}$.  The indices $x,y$ index the
spatial pixels (spaxels), and $z$ indexes the spectral layers of the
datacube. Mappings between $x,y$ and sky position (right ascension and
declination), as well as between $z$ and wavelength $\lambda$ can
be included in the metadata \citep{Greisen2002,Greisen2006}.  For the new
generation of wide-field IFUs based on image-slicers, the sky is
sampled contiguously, and typical dimensions of a MUSE datacube are
$x_\mathrm{max},y_\mathrm{max},z_\mathrm{max} \simeq 3 \times 10^2, 3
\times 10^2, 4 \times 10^3 $, that is, a datacube consists of
$\sim 4\times 10^8$ voxels.

We make a number of assumptions regarding the data structure, guided by the output
of the MUSE data reduction system \citep[DRS;][]{Weilbacher2012,Weilbacher2014}:
\begin{itemize}
\item We assume that atmospheric line and continuum emission has been 
subtracted from $\bm{F}$ \citep{Streicher2011,Soto2016}; see Sect.~\ref{sec:effective-variances} for a brief discussion on how to account for sky subtraction residuals within \texttt{LSDCat}. 
\item \texttt{LSDCat} currently requires a rectilinear grid in $x, y, z$.
In particular, the mapping between $z$ and $\lambda$ has to be linear with 
a fixed increment $\Delta\lambda$ per spectral layer, 
that is,
\begin{equation}
  \label{eq:4}
  \lambda = \lambda_{z=0} + z\,\Delta \lambda\;\text{,}
\end{equation}
where $\lambda_{z=0}$ designates the wavelength corresponding to the
wavelength of the first spectral layer.

For MUSE, this is achieved by the DRS through resampling the raw CCD data 
into the final datacube $\bm{F}$. We also demand a constant mapping
between $x,y$ and sky position for all spectral layers $z$, which the MUSE
pipeline accounts for in the resampling step by correcting for the wavelength-dependent
lateral offset (differential atmospheric refraction) along the parallactic angle 
\citep[e.g.][]{Filippenko1982,Roth2006}.  
\item Together with the flux datacube $\bm{F}$, \texttt{LSDCat} expects a second
cube  $\bm{\sigma}^2$ containing voxel-by-voxel variances. While such a 
variance cube is provided by the MUSE DRS as formal propagation of 
the various detector-level noise terms through the data reduction steps, 
the pipeline currently neglects the covariances between adjacent voxels 
introduced by the resampling process. We discuss this issue and some practical
considerations further in Sect.~\ref{sec:effective-variances}; here we simply 
assume that a datacube $\bm{\sigma}^2$ with appropriate variance estimates
is available.
\end{itemize}

As a very useful preparatory step for \texttt{LSDCat}, we recommend
that galaxies and stars with bright continuum emission (i.e. sources with
significant signal in the majority of spectral bins) are
subtracted from $\bm{F}$.  While the presence of such sources
does not render the detection and cataloguing algorithm unusable,
continuum bright sources may lead to (possibly many) catalogue
entries unrelated to actual emission-line objects.  We give some
guidance for the subtraction of continuum bright sources prior to 
running \texttt{LSDCat} in Sect.~\ref{sec:dealing-with-bright}.

Figure~\ref{fig:lsdcat_flow} depicts, as a flowchart, all the main
processing steps of \texttt{LSDCat}, leading from the input datacube
$\bm{F}$ and its associated variances $\bm{\sigma}^2$ to a catalogue
of emission lines.  Each processing step is implemented as a
stand-alone Python\footnote{Python Software Foundation. Python Language
  Reference, version 2.7. Available at \url{http://www.python.org}}
program.  The file format of the input data and variance cubes has to
conform to the FITS standard \citep{Pence2010}.  \texttt{LSDCat}
requires routines provided by NumPy\footnote{NumPy version 1.10.1.
  Available at \url{http://www.numpy.org/}.}  \citep{Walt2011},
SciPy\footnote{SciPy version 0.16.1, available at
  \url{http://scipy.org/}} \citep{scipy}, and Astropy\footnote{AstroPy
  version 1.0.1, available at \url{http://www.astropy.org/}}
\citep{AstropyCollaboration2013}.  For performing its operations,
\texttt{LSDCat} needs up to four datacubes loaded simultaneously.
Hence, for typical MUSE datacubes that contain $\sim 10^8 - 10^9$
32-bit floating point numbers, a computer with at least 16 GB of
random access memory is recommended.  Table~\ref{tab:lsdcat_routines}
lists the names of the individual \texttt{LSDCat} routines
implementing the various processing steps, together with the main
input parameters that govern the detection and cataloguing process.
In the following we describe each of the processing steps with its
routines in more detail.  A practical example of using the
\texttt{LSDCat} routines is provided in Appendix \ref{sec:usage-example}.

\subsection{3D matched filtering} 
\label{sec:3d-matched-filtering}

\begin{figure*}
  \centering
  \includegraphics[width=0.9\textwidth]{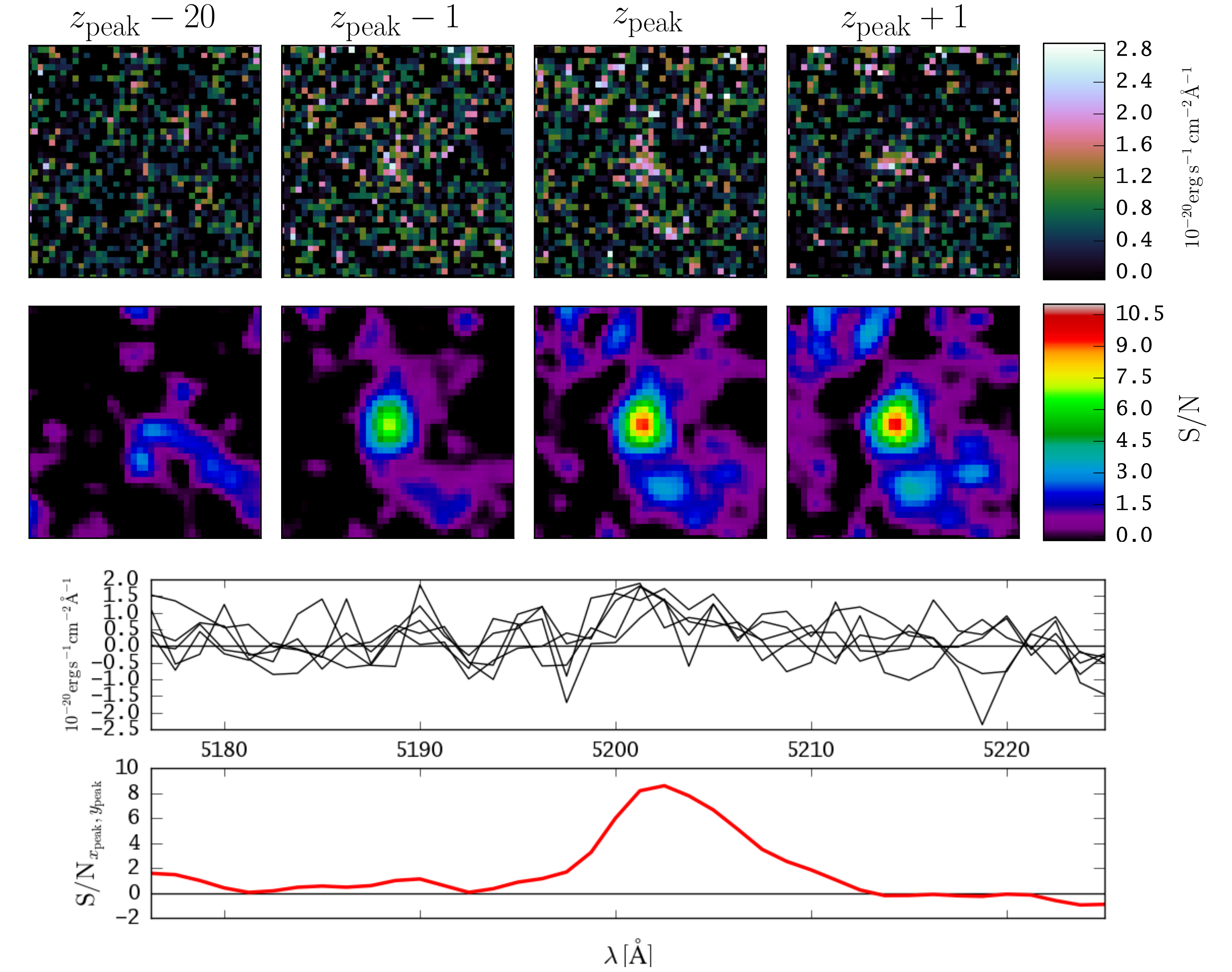}
  \caption{Example of the effect of matched filtering on the
    detectability of a faint emission-line source in the MUSE datacube
    of the Hubble Deep Field South \citep{Bacon2015}.  Shown is a
    Lyman $\alpha$ emitting galaxy at redshift $z=3.278$ with flux
    $F_\mathrm{Ly\alpha} = 1.3\times10^{-18}$\,erg\,s$^{-1}$ (ID \#162
    in \citealt{Bacon2015}). The panels in the first row display four
    different spectral layers of the continuum-subtracted datacube
    $\bm{F}$.  The second row shows the same spectral layers, but for
    the filtered $\bm{S/N}$ cube (Eq.~\ref{lsd_eq:12}) that was used to
    build the catalogue of emission line sources via thresholding
    (Eq.~\ref{lsd_eq:13}).  The leftmost panels show a layer
    significantly away from the emission line peak, the other panels
    show layers at spectral coordinates $z_{\mathrm{peak}}-1$,
    $z_{\mathrm{peak}}$, and $z_{\mathrm{peak}}+1$, respectively,
    where $z_{\mathrm{peak}}$ designates the layer containing the
    maximum $S/N$ value of the source.  The third row shows the flux
    density spectrum in the spaxels at
    ($x_\mathrm{peak},y_\mathrm{peak}$) and
    ($x_\mathrm{peak}\pm1, y_\mathrm{peak}\pm1$), where
    $x_\mathrm{peak}$ and $y_\mathrm{peak}$ are the spatial
    coordinates of the highest S/N value.  The bottom row shows a S/N
    spectrum extracted from the $\bm{S/N}$ cube at $x_\mathrm{peak}$
    and $y_\mathrm{peak}$.}
  \label{fig:detsnexample} 
\end{figure*}

The optimal detection statistic of an isolated signal in a dataset
with additive white Gaussian noise is given by the matched filter
transform of the dataset
\citep[e.g.][]{Schwartz1975,Das1991,Bertin2001,Zackay2015,Vio2016}.  This
transform cross-correlates the dataset with a template that matches the
properties of the signal to be detected.  

We utilise the matched filtering approach in \texttt{LSDCat} to
obtain a robust detection statistic for isolated emission line sources
in wide-field IFS datacubes. For a symmetric 3D template $\bm{T}$,
this is equivalent to a convolution of $\bm{F}$ with $\bm{T}$:
\begin{equation}
  \label{lsd_eq:1}
  \bm{\tilde{F}} = \bm{F} \ast \bm{T} \;\text{.}
\end{equation}
Here $\ast$ denotes the (discrete) convolution operation, that is, every voxel of
$\bm{\tilde{F}}$ is given by
\begin{equation}
  \label{lsd_eq:2}
  \tilde{F}_{x,y,z} = \sum\nolimits_{i,j,k} F_{i,j,k} \, T_{x-i,y-j,z-k} =
  \sum\nolimits_{i,j,k} T_{i,j,k} \, F_{x-i,y-j,z-k} \; \text{.}
\end{equation}
In principle, the summation runs over all dimensions of the
datacube, but in practice, terms where $T_{i,j,k} \approx 0$ can be
neglected.  Propagating the variances from $\bm{\sigma}^2$ through
Eq.~(\ref{lsd_eq:2}) yields the voxels of $\bm{\tilde{\sigma}}^2$:
\begin{equation}
  \label{lsd_eq:3}
  \tilde{\sigma}^2_{x,y,z} = \sum_{i,j,k} T_{i,j,k}^2 \,
  \sigma^2_{x-i,y-j,z-k} \; \text{.}
\end{equation}

The template $\bm{T}$ must be chosen such that its spectral and
spatial properties match those of a compact expected emission-line
source in $\bm{F}$.  \texttt{LSDCat} is primarily intended to search
for faint compact emission-line sources.  For such sources it is a
reasonable assumption that their spatial and spectral properties are
independent of one another.  We can therefore write $\bm{T}$ as a
product
\begin{equation}
  \label{lsd_eq:4}
  \bm{T} = \bm{T}^{\mathrm{spec}} \,\bm{T}^{\mathrm{spat}} \;\text{,}
\end{equation}
where $\bm{T}^{\mathrm{spat}}$ is the expected spatial profile and
$\bm{T}^{\mathrm{spec}}$ is the expected spectral profile.  The
voxels of $\bm{T}$ are thus given by
\begin{equation}
  \label{lsd_eq:5}
  T_{x,y,z} = T^{\mathrm{spat}}_{x,y} \, T^{\mathrm{spec}}_{z}\;\text{.}
\end{equation}
Of course, spatially extended sources with a velocity profile along
the line of sight are not optimally captured by separating the filter
into the spectral and spatial domain.  However, as we detail in
Sects.~\ref{sec:optim-choice-seeing}
and~\ref{sec:optim-choice-v_mathr}, moderate template mismatches do
not result in a major reduction of the maximum detectability that
could be achieved with an exactly matching template. 

Now with Eq.~(\ref{lsd_eq:5}), the 3D convolution of Eq. (\ref{lsd_eq:2}) yielding
$\bm{\tilde{F}}$ can be performed as a succession of two separate
convolutions, in no particular order: A spatial convolution with the
appropriate $\bm{T}^{\mathrm{spat}}$ in each spectral layer $z$ , and
a spectral convolution with $\bm{T}^{\mathrm{spec}}$ in each spaxel
$x,y$:
\begin{eqnarray}
  \label{lsd_eq:6}
  \tilde{F}_{x,y,z} &= \sum_{k}  T^{\mathrm{spec}}_{k} \left ( \sum_{i,j}
  T^{\mathrm{spat}}_{i,j}  \, F_{x-i,y-j,z-k} \right ) \phantom{\;\text{.}}\\ \label{lsd_eq:7}
  {} &= \sum_{i,j}
  T^{\mathrm{spat}}_{i,j}  \left ( \sum_{k}  T^{\mathrm{spec}}_{k} \,
       F_{x-i,y-j,z-k} \right ) \;\text{.}
\end{eqnarray}

Note that since every spectral layer $z$ of the datacube is convolved with a 
different spatial template, and since the width of the spectral template also changes 
with $z$ (see Sect.~\ref{sec:templates}), the equivalence between Eq.~(\ref{lsd_eq:6}) 
and Eq.~(\ref{lsd_eq:7}), mathematically speaking, is not strictly correct.
However, the variations of the templates with $\lambda$ are much slower
than the typical line widths of the spectral templates, meaning that  
we can approximate the convolution by using a {locally} invariant template.

As an indicator of the presence or absence of an emission line in $\bm{F}$
at a position $x,y,z$ we then evaluate the statistic
\begin{equation}
  \label{lsd_eq:12}
  \mathit{S/N}_{x,y,z} =
  \frac{\tilde{F}_{x,y,z}}{\tilde{\sigma}_{x,y,z}} \;\text{,}
\end{equation}
with $\tilde{F}_{x,y,z}$ from Eq.~(\ref{lsd_eq:2}) and
$\tilde{\sigma}_{x,y,z}$ as the square root of
Eq.~(\ref{lsd_eq:3}).  The voxels $\mathit{S/N}_{x,y,z}$ constitute
the signal-to-noise cube
\begin{equation}
  \label{lsd_eq:12a}
  \bm{\mathit{S/N}}\equiv\bm{\tilde{F}} / \bm{\tilde{\sigma}}
\;\text{,}\end{equation}
computed after the matched filtering has been performed.  
The values on the left side of Eq.~(\ref{lsd_eq:12})
can be translated into a probability for rejecting the null-hypothesis 
of no source being present at position $x,y,z$ in $\bm{F}$.  This is
commonly referred to as the detection significance of a source.
However, in a strict mathematical sense this direct translation is
only valid for sources that are exactly described by the
adopted template $\bm{T}$, in a dataset where the variance
terms on the right-hand side of Eq.~(\ref{lsd_eq:3}) are stationary and
fully uncorrelated.  Nevertheless, even if these strict requirements
are not met by the IFS datacube, filtering with $\bm{T}$ will always
reduce high-frequency noise while enhancing the presence of sources
that are similar in appearance with $\bm{T}$.  Thus
Eq.~(\ref{lsd_eq:12}) can still be used as a robust empirical measure
of the significance of a source present in $\bm{F}$ at a position $x,y,z$.
We illustrate this for a faint high-redshift Lyman $\alpha$-emitting galaxy
observed with MUSE in Fig.~\ref{fig:detsnexample}; this source is detected with 
high significance at a peak $S/N$ value of $\sim$9 in the 
$\bm{S/N}$-cube, although it is barely visible in the monochromatic 
layers of the original MUSE datacube.

In \texttt{LSDCat}, the spatial convolution is performed by the routine
\texttt{lsd\_cc\_spatial.py} and the spectral convolution is performed
by \texttt{lsd\_cc\_spectral.py}.  The output of a subsequent run of
those routines is a FITS file with two header and data units; one
storing $\bm{\tilde{F}}$ and the other one storing
$\bm{\tilde{\sigma}}^2$.  Both routines are capable of
fully leveraging multiple processor cores, if available, to process several datacube
layers or spaxels simultaneously.  Moreover, in
\texttt{lsd\_cc\_spatial.py} the computational time is reduced by using the fast Fourier transform
method provided by SciPy to convolve the individual
layers.  However, since the spectral kernel varies with wavelength we
cannot utilise the same approach in
\texttt{lsd\_cc\_spectral.py}, but  the discrete 1D convolution
operation can be written as a matrix-vector product with the
convolution kernel given by a sparse banded matrix
\citep[e.g.][Sect. 3.3.6]{Das1991}.  Hence here we achieve a substantial
acceleration of the computational speed by utilising 
the sparse matrix multiplication algorithm of SciPy.  Typical execution times of
\texttt{lsd\_cc\_spatial.py} and \texttt{lsd\_cc\_spectral.py} running in sequence
on a full MUSE $300\times300\times4000$ datacube, including reading
and writing of the data, are $\sim5$ minutes on an Intel Core-i7
workstation with 8 cores, or $\sim2$ minutes on a workstation with
two AMD Opteron 6376 processors with 32 cores each.

\subsection{Templates} 
\label{sec:templates} 

In line with the approximate separation of spatial and spectral filtering
described in the previous subsection, \texttt{LSDCat} requires two
templates, a spatial and a spectral one.

For the spatial template $\bm{T^{\mathrm{spat}}}$ the user has
currently the choice between a circular Gaussian profile,
\begin{equation}
  \label{lsd_eq:8}
  T_{x,y} = \frac{1}{2 \pi \sigma_\mathrm{G}^2} 
  \exp 
  \left ( 
    - \frac{x^2 + y^2}{2 \sigma_\mathrm{G}^2}
  \right )
\;\text{,}
\end{equation}
with the standard deviation $\sigma_\mathrm{G}$, or a
\cite{Moffat1969} profile,
\begin{equation}
  \label{lsd_eq:9}
  T_{x,y} = \frac{\beta - 1}{\pi r_\mathrm{d}^2} 
  \left [
    1 + \frac{x^2 + y^2}{r_\mathrm{d}^2}
  \right ]^{-\beta}
\;\text{,}
\end{equation}
with the width parameter $r_\mathrm{d}$ and the kurtosis parameter
$\beta$.  Both functions\footnote{In the limiting case
  $\beta \rightarrow \infty$ Eq.~(\ref{lsd_eq:9}) is equal to
  Eq.~(\ref{lsd_eq:8}); cf. \cite{Trujillo2001}.} are commonly used as
approximation of the seeing induced point spread function (PSF) in
ground-based optical and near-IR observations
\citep[e.g. ][]{Trujillo2001a,Trujillo2001} and therefore are well suited as 
spatial templates for compact emission line sources. Of course it is also 
possible to adopt a spatial template more extended than the PSF by 
choosing a correspondingly large value of $\sigma_\mathrm{G}$ or
$r_\mathrm{d}$. As discussed in Sects.~\ref{sec:optim-choice-seeing} 
and \ref{sec:optim-choice-v_mathr} below, the maximum $S/N$ delivered
by the matched filter behaves very benignly with respect to modest 
template mismatch, and the application of \texttt{LSDCat} is thus 
by no means limited to search for point sources.

Typically the parameter used to characterise the atmospheric seeing 
is the full width at half maximum (FWHM) of the PSF. 
For Eq.~(\ref{lsd_eq:8}) this is
\begin{equation}
\mathrm{FWHM}=2\,\sqrt{2 \ln 2}\, \sigma 
\label{eq:1}
\;\text{,}
\end{equation}
and for the Moffat in Eq.~(\ref{lsd_eq:9}) it is
\begin{equation}
\mathrm{FWHM}=2\,\sqrt{2^{1/\beta} - 1}\,r_\mathrm{d}\;\mathrm{.}
\label{eq:moffatfwhm}
\end{equation}
Generally, the seeing depends on wavelength.  Specifically, in MUSE
data and adopting the Moffat form to describe the PSF, the variation
of FWHM with $\lambda$ appears to be mainly driven by $r_\mathrm{d}$,
while $\beta$ appears to be close to constant
\citep[e.g.,][]{Husser2016}.  In \texttt{LSDCat} we approximate the
$\mathrm{FWHM}(\lambda)$ dependency as a quadratic polynomial
\begin{equation}
\mathrm{FWHM}(\lambda)[\text{\arcsec}] = p_0 + p_1(\lambda - \lambda_0) +
p_2 (\lambda - \lambda_0)^2
\label{eq:2}
\;\text{,}\end{equation}
 where the coefficients $p_0$[\arcsec],
$p_1$[\arcsec$/$\AA{}], $p_2$[\arcsec$/$\AA{}$^2$], 
and the reference wavelength $\lambda_0$ are input
parameters supplied by the user. In Sect.~\ref{sec:optim-choice-seeing}
we further discuss the adopted PSF parametrisation and the choice of 
suitable parameter values.

As a spectral template in \texttt{LSDCat} we employ a simple 1D Gaussian
\begin{equation}
\label{lsd_eq:10}
  T_z = 
  \frac{1}{\sqrt{2\pi}\sigma_z}
  \exp \left 
    ( - \frac{z^2}{2\sigma_z^2}
  \right ) \; \text{,}
\end{equation}
with the standard deviation $\sigma_z$. While the line profiles of 
actual galaxies may deviate in detail from this idealised function, 
the deviations are usually minor and do not lead to significant
changes in the $S/N$ achieved by the matched filter when compared
to a simple Gaussian. 

Since velocity broadening usually dominated the widths of emission lines 
from galaxies, \texttt{LSDCat} assumes the width of the spectral 
template to be fixed in velocity space, $\sigma_v = \mathrm{const}$.
As long as the mapping between $\lambda$ and spectral coordinate $z$ 
is linear, $\sigma_z$ in Eq.~(\ref{lsd_eq:10})
depends linearly on $z$ when parameterised by $\sigma_v$:
\begin{equation}
  \label{lsd_eq:11}
  \sigma_z
  = 
  \frac{\sigma_v}{c}
  \left (
    \frac{\lambda_{z=0}}{\Delta \lambda} + z 
  \right ) \; \text{.}
\end{equation}
The input parameter supplied by the \texttt{LSDCat} user is the velocity FWHM
of the Gaussian profile $\mathrm{FWHM}_v = 2 \sqrt{2\ln 2}\,\sigma_v$.
In Sect.~\ref{sec:optim-choice-v_mathr} we show that the $S/N$ of any
given emission line does not depend very sensitively on the chosen value
of $\mathrm{FWHM}_v$, and in many cases a single spectral template with
a typical value of $\mathrm{FWHM}_v$ may be sufficient.

\subsection{Thresholding}
\label{sec:thresh-gener-interm}

A catalogue of emission line sources can now be constructed by 
thresholding the $\bm{S/N}$-cube with a user-specified value 
$\mathit{S/N}_\mathrm{det.}$.  This threshold is used to create a binary cube $\bm{L}$ 
with voxels given by
\begin{equation}
  \label{lsd_eq:13}
  L_{x,y,z} = 
  \begin{cases}
    1&\mbox{if } \mathit{S/N}_{x,y,z} \geq
    \mathit{S/N}_\mathrm{det}\; \text{,} \\
    0&\mbox{otherwise.}
  \end{cases}
\end{equation}
The detection threshold $\mathit{S/N}_\mathrm{det}$ is the principal
input parameter to be set by the \texttt{LSDCat} user.  Each cluster
of non-zero neighbouring voxels in $\bm{L}$ (6-connected topology)
constitutes a detection.  A high threshold will lead to a small number of
highly significant detections, while lower values of $\mathit{S/N}_\mathrm{det}$
will lead to more entries in the catalogue, but also increase the chance of
including entries that are spurious, that is, that do not correspond to real 
emission line objects. We give guidelines on the choice of the detection 
threshold based on our experience with
MUSE data in Sect.~\ref{sec:gudel-optim-param} below.

For each detection, \texttt{LSDCat} records the coordinates
$x_\mathrm{peak}$, $y_\mathrm{peak}$,$z_\mathrm{peak}$ of the local
maximum in the $\bm{S/N}$-cube, its value
$\mathit{S/N}_\mathrm{peak}\equiv\mathit{S/N}_{x_\mathrm{peak},y_\mathrm{peak},z_\mathrm{peak}}$,
and the number of voxels $N_\mathrm{det}$ constituting the detection.
These values constitute an intermediate catalogue of detections.  Each
entry is assigned a unique running integer number $i$.  Moreover, \texttt{LSDCat} also assigns an integer
\emph{object} identifier $j_\mathrm{Obj}$ to detection clusters
occurring at different wavelengths but similar spatial positions within
a small search radius to account for sources with multiple
significantly detectable emission lines.  This intermediate catalogue
is created by the routine \texttt{lsd\_cat.py}, utilising routines
from the SciPy \texttt{ndimage.measurements} package.  The
intermediate catalogue is written to disk as a FITS binary table.  The
actual execution time is generally much shorter than the previous step
of applying the matched filter, but it depends on the detection
threshold that determines the number of objects.

\subsection{Measurements}
\label{sec:meas}

\begin{table*}
  \caption{Output parameters for each \texttt{LSDCat}
    detection.}
  \label{tab:params}
  \centering
  \begin{tabular}{lll} \hline\noalign{\smallskip} 
    Output Parameter(s) & \texttt{LSDCat} Name & Description \\ 
    \noalign{\smallskip}\hline\noalign{\smallskip}
    $i$                 & \texttt{I}           & Running ID; see Sect.~\ref{sec:thresh-gener-interm} \\ 
    $j_\mathrm{Obj} $   & \texttt{ID}          & Object identifier; see Sect.~\ref{sec:thresh-gener-interm} \\
    $x_\mathrm{peak}$, $y_\mathrm{peak}$, $z_\mathrm{peak}$ & \texttt{\{X,Y,Z\}\_PEAK\_SN} & $\mathit{SN}_\mathrm{peak}$ coordinate; see  Sect.~\ref{sec:thresh-gener-interm} \\ 
    $N_\mathrm{pix}$   & \texttt{NPIX} & Number of voxels above $\mathit{S/N}_\mathrm{det}$; see Sect.~\ref{sec:thresh-gener-interm} \\
    $\mathit{S/N}_\mathrm{peak}$ & \texttt{DETSN\_MAX} & $S/N$ value at ${x_\mathrm{peak},y_\mathrm{peak},z_\mathrm{peak}}$; see Sect.~\ref{sec:thresh-gener-interm} \\[0.5ex]
    $x_{S/N}^\mathrm{com}$, $y_{S/N}^\mathrm{com}$, $z_{S/N}^\mathrm{com}$ &  \texttt{\{X,Y,Z\}\_SN} & $S/N$-weighted centroid; Eq.~(\ref{lsd_eq:14})\\[0.5ex]
    $x_{F}^\mathrm{com}$, $y_{F}^\mathrm{com}$, $z_{F}^\mathrm{com}$ &  \texttt{\{X,Y,Z\}\_FLUX} & $F$-weighted centroid; Eq.~(\ref{lsd_eq:14}) with $\mathit{S/N}_{x,y,z}$ substituted by $F_{x,y,z}$ \\[0.5ex]
    $x_{\tilde{F}}^\mathrm{com}$, $y_{\tilde{F}}^\mathrm{com}$, $z_{\tilde{F}}^\mathrm{com}$ &  \texttt{\{X,Y,Z\}\_SFLUX} & $\tilde{F}$-weighted centroid; Eq.~(\ref{lsd_eq:14}) with $\mathit{S/N}_{x,y,z}$ substituted by $\tilde{F}_{x,y,z}$ from Eq.~(\ref{lsd_eq:2}) \\[0.5ex]
    $z_\mathrm{min}^{\mathrm{NB}}$, $z_\mathrm{max}^{\mathrm{NB}}$ & \texttt{Z\_NB\_\{MIN,MAX\}} & Minimum or maximum $z$ coordinate above  $\mathit{S/N}_\mathrm{ana}$ in $\bm{S/N}$-cube \\[0.5ex]
    $x^{(1)},y^{(1)}$ & \texttt{\{X,Y\}\_1MOM}  &  2D moment-based centroid in NB$_{x,y}(\tilde{F})$ image (Eq.~\ref{lsd_eq:15}, Eq.~\ref{lsd_eq:16})  \\
    $x^{(2)},y^{(2)},{[xy]}^{(2)}$ & \texttt{\{X,Y,XY\}\_2MOM} & 2D second central moments in NB$_{x,y}(\tilde{F})$ image (Eq.~\ref{lsd_eq:17}, Eq.~\ref{lsd_eq:18}, Eq.~\ref{lsd_eq:19}) \\
    $R_\sigma$ & \texttt{R\_SIGMA} & $R_\sigma = \sqrt{ (x^{(2)} + y^{(2)})/2 }$ \\
    $R_\mathrm{Kron}$ & \texttt{R\_KRON} & Kron radius (Eq.~\ref{lsd_eq:17}) \\
    $F(k\cdot R_\mathrm{Kron})$ & \texttt{FLUX\_\{}$k$\texttt{\}KRON} & Integrated flux in $k\times R_\mathrm{Kron}$ aperture (Eq.~\ref{lsd_eq:21})\\
    $\sigma_F (k\cdot R_\mathrm{Kron})$ & \texttt{ERR\_FLUX\_\{}$k$\texttt{\}KRON} & Uncertainty on $F(k\times R_\mathrm{Kron})$ \\
    \noalign{\smallskip}\hline
  \end{tabular}
  \tablefoot{
    A comma separated list within brackets in the second column
    indicates the set of corresponding \texttt{LSDCat} output column names.
    For each $x,y$ coordinate there is also a corresponding right
    ascension and declination value available and for each $z$
    coordinate a corresponding wavelength can be tabulated if a
    corresponding world coordinate system is
    specified.}
\end{table*}

\texttt{LSDCat} provides a set of basic measurement parameters for
each detection of the intermediate catalogue.  These parameters are
determined using the datacubes $\bm{F}$, $\bm{\sigma}^2$,
$\bm{\tilde{F}}$ and $\bm{S/N}$.  The set of parameters is chosen to be
robust and independent from a specific scientific application.  For
more complex measurements involving, for example, fitting of flux distributions, 
the \texttt{LSDCat} measurement capability can serve as a starting point.

In Table~\ref{tab:params} we list the various output
parameters that are generated for each detection. 
This parametrisation of the detected sources and their emission lines
constitutes the final processing step of \texttt{LSDCat}.
The routine \texttt{lsd\_cat\_measure.py} performs the parameterisation task
and writes out the final catalogue.  Running \texttt{lsd\_cat\_measure.py} 
on an intermediate catalogue with $\sim10^2$ entries takes typically 100 
seconds on the two machines mentioned above in Sect.2.2, with most of the
time spent on reading the four input datacubes.

\subsubsection{Centroids}
\label{sec:centroids}

The coordinates of each detection local maximum in the
$\bm{S/N}$-cube (Sect.~\ref{sec:thresh-gener-interm}) serve only
as a first approximation of its spatial and spectral position.  As a
refinement \texttt{lsd\_cat\_measure.py} can calculate several
different sets of centroid positions for each detected line
cluster. For example, the 3D $S/N$-weighted centroid is given by the
first moments
\begin{multline}
  \label{lsd_eq:14}
  \left (x_{S/N}^\mathrm{com}, y_{S/N}^\mathrm{com}, z_{S/N}^\mathrm{com} \right)
  =  \\ 
  \left ( \,
    \frac{\sum_{x,y,z} x\cdot\mathit{S/N}_{x,y,z}  }{\sum_{x,y,z} \mathit{S/N}_{x,y,z}} \, , \,
    \frac{\sum_{x,y,z} y\cdot\mathit{S/N}_{x,y,z}  }{\sum_{x,y,z} \mathit{S/N}_{x,y,z}} \, , \,
    \frac{\sum_{x,y,z} z\cdot\mathit{S/N}_{x,y,z}  }{\sum_{x,y,z} \mathit{S/N}_{x,y,z}} 
  \, \right  ) \; \text{.}
\end{multline}
Here, for each detection, the summation runs over all non-zero
($x_\mathrm{peak}$, $y_\mathrm{peak}$,$z_\mathrm{peak}$)-neighbouring
voxels in a thresholded datacube similar to $\bm{L}$
(Eq.~\ref{lsd_eq:13}), but with voxels set to one if they are above an
\emph{analysis threshold} $\mathit{S/N}_\mathrm{ana}$.  This
additional threshold, which must be smaller or equal to
$\mathit{S/N}_\mathrm{det}$, is the required input parameter for
\texttt{lsd\_cat\_measure.py}.  Guidelines for choosing
$\mathit{S/N}_\mathrm{ana}$ are discussed in
Sect.~\ref{sec:optim-choice-detect}.  Similarly, \texttt{LSDCat} can
measure 3D centroid coordinates with the original flux cube $\bm{F}$
and with the filtered flux cube $\bm{\tilde{F}}$ as weights. In these
cases, Eq.~(\ref{lsd_eq:14}) is applied again, but with
$\mathit{S/N}_{x,y,z}$ being substituted by $F_{x,y,z}$ and
$\tilde{F}_{x,y,z}$, respectively.

3D centroids provide a non-parametric way of calculating spatial and
spectral positions, making use of the full 3D information present in
the datacube.  While the calculation using the flux cube is unbiased
against a particular choice of filter template, it is not very
robust for low $\mathit{S/N}$ detections.  This shortcoming is
alleviated for the centroids calculated on the filtered flux cube or
the $\bm{S/N}$-cube.  The latter, however, could potentially
be biased by local noise extrema, for example, at spectral layers near
sky-lines.

As an alternative approach to 3D centroids, \texttt{LSDCat} can
calculate 2D centroids on synthesised narrow-band images
$\mathrm{NB}_{x,y}(\tilde{F})$ from the filtered flux cube
$\bm{\tilde{F}}$:
\begin{equation}
  \label{lsd_eq:15}
   \mathrm{NB}_{x,y}(\tilde{F})
   =
   \sum\nolimits_{z=z_\mathrm{min}^{\mathrm{NB}}}^{z_\mathrm{max}^{\mathrm{NB}}} \tilde{F}_{x,y,z} 
   \; \text{.}
\end{equation}
The boundary indices of each synthetic narrowband image
$z_\mathrm{min}^{\mathrm{NB}}$ and $z_\mathrm{max}^{\mathrm{NB}}$ are
taken as the minimum and maximum $z$ coordinate of all voxels of a
detection above the analysis threshold $\mathit{S/N}_\mathrm{ana}$.
Then the 2D weighted centroid coordinates follows from the first image
moments:
\begin{equation}
  \label{lsd_eq:16}
  (x^{(1)},y^{(1)}) = 
  \left
    ( \frac{\sum_{x,y} x \cdot \mathrm{NB}_{x,y}(\tilde{F})}{\sum_{x,y}  \mathrm{NB}_{x,y}(\tilde{F})},
    \frac{\sum_{x,y} y \cdot \mathrm{NB}_{x,y}(\tilde{F})}{\sum_{x,y}  \mathrm{NB}_{x,y}(\tilde{F})}
  \right )
  \; \mathrm{.}
\end{equation}
In this equation the summation runs over all pixels $x,y$ of
$\mathrm{NB}_{x,y}(\tilde{F})$ that belong to the detection cluster and
are above the analysis threshold $\mathit{S/N}_\mathrm{ana}$ in the
$z_\mathrm{peak}$ layer of the $\bf{S/N}$-cube. 

The 2D narrowband images are furthermore used by \texttt{lsd\_cat\_measure.py} 
to derive basic shape information using the second central image
moments of each detection. These are defined as
\begin{align}
  \label{lsd_eq:17}
  x^{(2)} =& \frac{\sum_{x,y} x^2 \cdot \mathrm{NB}_{x,y}(\tilde{F})}{\sum_{x,y}  \mathrm{NB}_{x,y}(\tilde{F})} - (x^{(1)})^2 \;\text{,}\\
  \label{lsd_eq:18}
  y^{(2)} =& \frac{\sum_{x,y} y^2 \cdot \mathrm{NB}_{x,y}(\tilde{F})}{\sum_{x,y}  \mathrm{NB}_{x,y}(\tilde{F})} - (y^{(1)})^2 \;\text{,}\\
\label{lsd_eq:19}
  {[xy]}^{(2)} =& \frac{\sum_{x,y} x^2 \cdot y^2 \cdot \mathrm{NB}_{x,y}(\tilde{F})}{\sum_{x,y}  \mathrm{NB}_{x,y}(\tilde{F})} - x^{(1)} \cdot y^{(1)} \; \text{.}
\end{align}
These values are simple indicators for the spatial extent and
elongation of a detected emission line cluster. For example,
${[xy]}^{(2)}\equiv 0$ for a circular symmetric distribution of the
object in $\mathrm{NB}_{x,y}(\tilde{F})$.  Moreover, in this case the radius
\begin{equation}
  \label{eq:Rsigma}
  R_\sigma = \sqrt{ (x^{(2)} + y^{(2)})/2 }  
\;\text{,}
\end{equation}
encircles 68\% of the flux in the filtered narrowband image for a
perfect point source blurred by a Gaussian PSF.

While in principle, the original flux datacube $\bm{F}$ could also be
used in Eqs.~(\ref{lsd_eq:15}) to (\ref{lsd_eq:19}), in practice, the
calculation of the moments directly from the flux cube is relatively susceptible to noise for very faint sources.  Our experience with MUSE
data showed that the centroids and shapes determined in the 2D
narrowband images based on the filtered datacubes provide the most
reliable measurements even for low $\mathit{S/N}$ detections.
Moreover, the spatial coordinates from Eq.~(\ref{lsd_eq:16}) are in
closest agreement with the centroids determined in broad-band imaging
data.

\subsubsection{Integrated line fluxes}
\label{sec:integr-line-flux}

The difficulty in measuring the total flux of a detected line lies in its 
unknown spectral shape and in the unknown spatial distribution of the flux.
While for the {detection} it is not required to accurately know 
these properties (as long as the template mismatch is not too poor),
the template scaling factor depends very critically on the degree of similarity
between source and template and can therefore not be used as 
flux indicator. This is different from PSF matching techniques in 
stellar fields \citep[e.g.][]{Kamann2013}, 
where all objects are point sources and no strong mismatches are expected. 
The task for \texttt{LSDCat} is different: We have to define 3D boundaries for 
each emission line  to allow for the summation of voxel values 
within these boundaries as flux measurement, but avoid the inclusion
of too many unrelated empty-sky voxels that would compromise the
precision of the measurement. In line with the prime purpose of 
\texttt{LSDCat} as a {detection} tool, we implemented an approach
that emphasises robustness over sophistication.

\texttt{LSDCat} addresses this task in two steps. The first is the construction
of a narrowband image via setting an analysis $S/N$ threshold as described 
in the previous subsection. A reasonable choice of $\mathit{S/N}_\mathrm{ana}$ 
should produce spectral boundaries $z_\mathrm{min}^{\mathrm{NB}}$ and $z_\mathrm{max}^{\mathrm{NB}}$ that enclose the emission line (more or less) 
completely in the original datacube $\bm{F}$.  Therefore the
pixels of the narrow-band image $\mathrm{NB}_{x,y}(F)$ created
by summation of $\bm{F}$ from $z_\mathrm{min}^{\mathrm{NB}}$ to
$z_\mathrm{max}^{\mathrm{NB}}$ (analogous to
Eq.~\ref{lsd_eq:15}) can be used for flux integration. 

The second step is then to derive suitable apertures for these narrowband
images, taking the spatial extent of each source into account. Currently, 
\texttt{LSDCat} measures fluxes in circular apertures with radii
defined as multiples $k$ of the Kron radius \citep{Kron1980} of a detected line. 
The Kron radius is however determined not in the measured dataset itself, 
but again in the narrowband image based on the {filtered} datacube, 
which makes the procedure much more robust especially at low S/N. 
The measured quantity is then $F(k \cdot R_\mathrm{Kron})$ with
\begin{equation}
  \label{lsd_eq:20}
  R_\mathrm{Kron}
  =
  \frac{\sum_{x,y} \mathrm{NB}_{x,y}(\tilde{F}) \sqrt{(x-x^{(1)})^2 + (y-y^{(1)})^2}}{\sum_{x,y} \mathrm{NB}_{x,y}(\tilde{F})} \; \text{.}
\end{equation}
In order to avoid possibly unphysically small or large values of $R_\mathrm{Kron}$ 
caused by artefacts in the data, a minimal and a maximal value for $R_\mathrm{Kron}$ 
can be set by the user.

The factor $k$ is also defined by the \texttt{LSDCat} user. Multiple
values of $k$ result in multiple columns of the output catalogue.  
The line flux $F_\mathrm{line}(k \cdot R_\mathrm{Kron})$ is then given by the sum
\begin{equation}
  \label{lsd_eq:21}
  F_\mathrm{line}(k \cdot R_\mathrm{Kron}) = \Delta\lambda\,\sum\nolimits_{x,y} \; \sum\nolimits_{z=z_\mathrm{min}^{\mathrm{NB}}}^{z_\mathrm{max}^{\mathrm{NB}}} F_{x,y,z} 
\;\text{,}\end{equation}
with the first sum running over all $x,y$ that satisfy
$\sqrt{(x-x^{(1)})^2 + (y-y^{(1)})^2} \leq R_\mathrm{Kron}$. The
aperture thus has cylindrical shape, with its symmetry axis going
through the 2D centroid position.  The factor $\Delta\lambda$
that denotes the increment per spectral layer is needed to convert the sum of voxel
values into a proper integral over the line.

It can be shown that the $2.5\,R_\mathrm{Kron}$ aperture includes
$\gtrsim$ 90\% of the total flux even for extended sources with relatively
shallow profiles, as long as the determination of the Kron radius in
Eq.~(\ref{lsd_eq:20}) accounts for pixels at sufficiently large radii
\citep[e.g.][]{Graham2005}.  We follow a similar approach as adopted
in \texttt{SExtractor} \citep{Bertin1996} by summing over all $x,y$ in
Eq.~(\ref{lsd_eq:20}) that satisfy
\begin{equation}
  \label{eq:11}
  \sqrt{(x-x^{(1)})^2 + (y-y^{(1)})^2} \leq 6 \times R_\sigma \;\text{,}
\end{equation}
with $R_\sigma$ as defined in Eq.~(\ref{eq:Rsigma}). The uncertainty
$\sigma_F$ of this flux measurement is obtained by propagating the
voxel variances $\sigma^2_{x,y,z}$ through Eq.~(\ref{lsd_eq:21}).

\section{Validation of the software}
\label{sec:validation-software}

\subsection{Creation of test datacubes}
\label{sec:creat-test-datac}

We now validate the correctness of the algorithms implemented in the
\texttt{LSDCat} software.  To this aim we produced a set of datacubes
that contain fake emission line sources at known positions with known
fluxes and extents.  Instead of utilising a completely artificial data
set with ideal noise, we based our source insertion experiment on the
MUSE HDFS datacube\footnote{MUSE HDFS version 1.24, available for
  download from \url{http://muse-vlt.eu/science/hdfs-v1-0/}}
\citep{Bacon2015}.  Thereby we ensure that our test data is identical
in noise (and potential systematics) with real
observations. Furthermore, we self-calibrated the noise by calculating
empirical variances as we recommend in
Sect.~\ref{sec:effective-variances}.

We implanted the fake emission line sources into a continuum-subtracted version of the HDF-S datacube at a wavelength of
5000\,\AA{}.  Continuum subtraction was performed with the
median-filter subtraction method detailed in
Sect.~\ref{sec:dealing-with-bright}.  The chosen insertion wavelength
ensures a clean test environment that is not hampered by systematic
sky-subtraction residuals which exist in the redder parts of the
datacube.  In total we created 23 test datacubes for the fake source
emission line fluxes $\log F\,[\mathrm{erg\,s^{-1}cm^{-2}}] = - 16.2$
to $-18.5$ in steps of $0.1$dex.  Each cube contains 51 fake sources
with the same emission line flux.  The spatial positions of the
implanted sources are based on the pseudo-random Sobol sequence
\citep[][, Sect. 7.7]{Press1992}.  The pseudo-random grid guarantees
that all sources have different distances to the edges of the
rectangular grid of MUSE slicer-stacks, thus possible systematic
effects from localised noise properties within this slicer-stack grid
are mitigated.  As test sources we utilised Gaussian emission lines
with a line width (FWHM) of 250\,km\,s$^{-1}$.  The sources were
assumed to be PSF-like and we approximated the PSF blurring by a 2D
Gaussian with 0.88\arcsec{} FWHM at 5000\,\AA{}, a value we obtained
from a 2D Gaussian fit to the brightest star in the HDFS field.  The
datacubes containing the fake sources as well as all processing steps
needed to reproduce the results from the validation exercise presented
in this section are available via the \texttt{LSDCat} software
repository.

\subsection{Minimum detectable emission line flux at a given
  detection threshold}
\label{sec:detection-threshold}

\begin{figure}
  \centering
  \resizebox{\hsize}{!}{\includegraphics{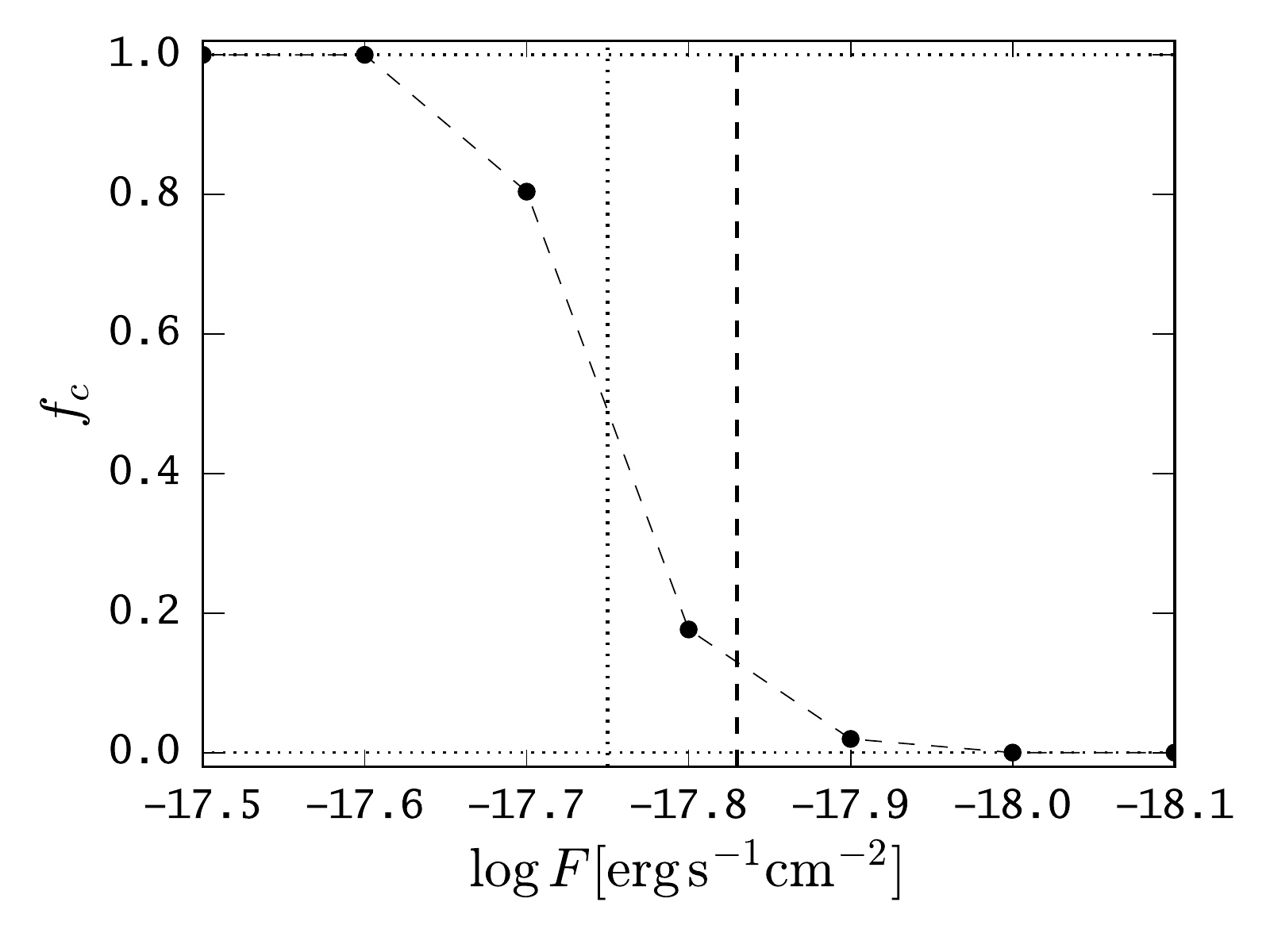}}
  \caption{Completeness curve
    $f_{\mathrm{C}}(\log F [\mathrm{erg\,s^{-1}cm^{-2}\AA{}^{-1}}])$
    from a fake source insertion and recovery experiment on the MUSE
    HDF-S datacube at 5000\,\AA{}.  The dotted vertical line shows the
    50\% completeness limit at
    $\log F [\mathrm{erg\,s^{-1}cm^{-2}}])=-17.75$ and the dashed
    vertical line shows the analytically approximated minimum line
    flux given in Eq.~(\ref{eq:10}) as
    $\log F_\mathrm{line} \;[\mathrm{erg\,s^{-1}\,cm^{-2}}] \approx
    -17.83$ at which \texttt{LSDCat} is expected to detect emission
    line sources in this experiment. }
\label{fig:compcurve}
\end{figure}

For known background noise, the minimum detection significance at which
an emission line can be recovered from the datacube is intrinsically
linked to the total flux of the line, its spatial and spectral
morphology, and the degree of mismatch between filter template and
emission line source signal.  The latter we discuss in 
Sects.~\ref{sec:optim-choice-seeing} and
\ref{sec:optim-choice-v_mathr} for the spectral and spatial filtering
processes, respectively, and Gaussian emission line expressions
for the detection significance attenuation due to shape mismatch are
presented in Eq.~(\ref{eq:zackay}) and Eq.~(\ref{eq:lsd_own}).

For the test datacubes described above we have complete control over
the shape parameters. Thus we can use the exact template in the matched
filtering process.  As a benchmark we will now derive an analytic
approximation for the minimum recoverable line flux at a given
detection threshold and then compare the result to a source recovery
experiment performed with \texttt{LSDCat} on the test datacube.

For the match between emission line and filter being perfect in the
datacube, the flux distribution of that emission line at position
$x',y',z'$ in the datacube can be written as
\begin{equation}
  \label{eq:3}
  F_{x,y,z} = \frac{F_\mathrm{line}}{\Delta \lambda}
  T^{\mathrm{spat}}_{x-x',y-y'} T^{\mathrm{spec}}_{z-z'}\;\text{,}
\end{equation}
with $F_\mathrm{line}$ being the total flux in that line,
$\Delta \lambda$ being the wavelength increment per spectral layer
defined in Eq.~(\ref{eq:4}), and $T^{\mathrm{spec}}_{z}$ and
$T^{\mathrm{spat}}_{x,y}$ being the spectral and spatial templates
given by Eq.~(\ref{lsd_eq:10}) and Eq.~(\ref{lsd_eq:8}) for the 1D and
2D Gaussian, respectively.  With Eq.~(\ref{eq:3}) and
Eq.~(\ref{lsd_eq:7}), we can thus write for the peak value of the
matched filtered datacube at position $x',y',z'$:
\begin{equation}
  \label{eq:5}
  \tilde{F}_{x',y',z'} =  \frac{F_\mathrm{line}}{\Delta \lambda}
  \sum_{i,j,k} \left ( T^{\mathrm{spat}}_{i,j} \right )^2  \left (
    T^{\mathrm{spec}}_{k} \right )^2 \; \text{.}
\end{equation}

Since the noise is usually not constant over the shape of the filter,
there exists no general solution for the error propagation given in
Eq.~(\ref{lsd_eq:3}).  To obtain an approximate solution, we
approximate $\sigma_{x,y,z}$ as being, on average, constant spectrally
and spatially, at least over all voxels within the matched filter:
$\sigma_{x,y,z} \approx \overline{\sigma}$.  Due to the absence of
strong sky emission lines in the blue part of the MUSE datacube this
approximation is well justified for the test data set that we consider
here.  Therefore Eq.~(\ref{lsd_eq:3}) can be written as
\begin{equation}
  \label{eq:6}
  \tilde{\sigma}^2_{x,y,z} \approx \overline{\sigma}^2 \sum_{i,j,k}
  \left ( T^{\mathrm{spat}}_{i,j} \right )^2  \left (
    T^{\mathrm{spec}}_{k} \right )^2 \; \text{.}
\end{equation}
Assuming that the dispersion $\sigma_Z$ and $\sigma_G$ of the 1D and
2D Gaussian in Eq.~(\ref{lsd_eq:8}) and Eq.~(\ref{lsd_eq:16}) are
large enough to make sampling and aliasing effects of the profiles in
the datacube negligible we can replace the sum with an integral, thus
\begin{equation}
  \label{eq:7}
   \sum_k \left ( T^{\mathrm{spec}}_{k} \right )^2 \approx
 \int_{-\infty}^{\infty} \frac{1}{2\pi\sigma_z^2}
  \exp \left 
    ( - \frac{z^2}{\sigma_z^2}
  \right )\, \mathrm{d}z = \frac{1}{2\sqrt{\pi}\sigma_z}
\;\text{,}
\end{equation}
and
\begin{equation}
  \label{eq:8}
  \sum_{i,j} \left ( T^{\mathrm{spat}}_{i,j}  \right )^2 \approx
  \iint_{-\infty}^{\infty} \frac{1}{4 \pi^2 \sigma_\mathrm{G}^4} 
  \exp 
  \left ( 
    - \frac{x^2 + y^2}{ \sigma_\mathrm{G}^2} 
  \right )\, \mathrm{d}x\mathrm{d}y = \frac{1}{4 \pi
    \sigma_\mathrm{G}^2}  \;\text{.}
\end{equation}
With these expressions in Eq.~(\ref{eq:5}) and Eq.~(\ref{eq:6}) we can
write the peak detection significance of the
line at position $x',y',z'$ via Eq.~(\ref{lsd_eq:12}) as
\begin{equation}
  \label{eq:9}
  \mathit{S/N}_{x',y',z'} \approx \frac{1}{\sqrt{8 \pi^{3/2} \sigma_G
      \sigma_z}} \times \frac{F_\mathrm{line}}{\overline{\sigma}
    \Delta \lambda}  \; \text{.}
\end{equation}

With the expression given in Eq.~(\ref{eq:9}) it is now possible to
estimate the minimum recoverable line flux at a given detection
threshold.  For the artificial sources implanted in the MUSE HDFS data
we have $\sigma_G=0.88\arcsec=1.84\,$px.  The spectral line width
$v_\mathrm{FWHM} = 250$\,km\,s$^{-1}$ translates to
$\sigma_z = 1.46$\,px at 5000\,\AA{} and $\Delta \lambda = 1.2$\AA{}.
By averaging our empirical noise estimate around 5000\,\AA{} we find
$\overline{\sigma}=1.42\times
10^{-20}$\,erg\,s$^{-1}$cm$^{-2}$\AA{}$^{-1}$. As we detail in
Sect.~\ref{sec:optim-choice-detect} for the MUSE HDFS datacube, a
detection threshold of $S/N_\mathrm{det} \approx 8$ is a value found
to be suitable for practical work. With the stated values inserted
into Eq.~(\ref{eq:9}) we calculate
\begin{equation}
  \label{eq:10}
  \log F_\mathrm{line} \;[\mathrm{erg\,s^{-1}\,cm^{-2}}] \approx -17.83
\end{equation}
as the minimum line flux at which sources should be detectable at
$S/N_\mathrm{det} = 8$, if an exactly matching filter was chosen in
the cross-correlation process.

We now perform the recovery experiment utilising the test datacubes
introduced in Sect.~\ref{sec:creat-test-datac} to check whether
\texttt{LSDCat} is indeed able to detect emission line sources with
$S/N_\mathrm{det} = 8$ at the estimated minimum line flux.  Therefore
we process all 23 test datacubes with \texttt{LSDCat} utilising the
perfect matched filter, as well as setting $S/N_\mathrm{det} = 8$.  In
the resulting catalogues we then search for positional cross-matches
with the input source positions.  From counting these cross-matches we
produce the completeness curve
$f_{\mathrm{C}}(\log F [\mathrm{erg\,s^{-1}cm^{-2}\AA{}^{-1}}])$
displayed in Fig.~\ref{fig:compcurve}.  This curve displays the
fraction of recovered sources as a function of input emission line
flux.  As vertical dashed and dotted lines, respectively, we show in
this figure the analytically approximated minimum flux of
Eq.~(\ref{eq:10}) for detectability and 50\% completeness estimate.

As can be seen in Fig.~\ref{fig:compcurve}, the completeness curve
starts to rise at the analytically estimated minimum flux for
detectability.  This validates the implementation of the detection
algorithm in \texttt{LSDCat}.  Nevertheless, it is also clear from
Fig.~\ref{fig:compcurve} that the rise of the completeness curve is
not from 0 to 1 at the estimated minimum, but it takes approximately 0.4\,dex
to recover all emission line sources at
$\log F [\mathrm{erg\,s^{-1}cm^{-2}\AA{}^{-1}}]=-17.6$; slightly above
the estimated minimum value.  Overall, however, there is an excellent match
between a simple analytic model of detectability and our realistic
implementation of a detection experiment, especially given the fact
that the noise in the real data is certainly not exactly Gaussian as
assumed in the model.

\subsection{Line flux measurements}
\label{sec:line-flux-meas}

\begin{figure}
  \centering
  \resizebox{\hsize}{!}{\includegraphics{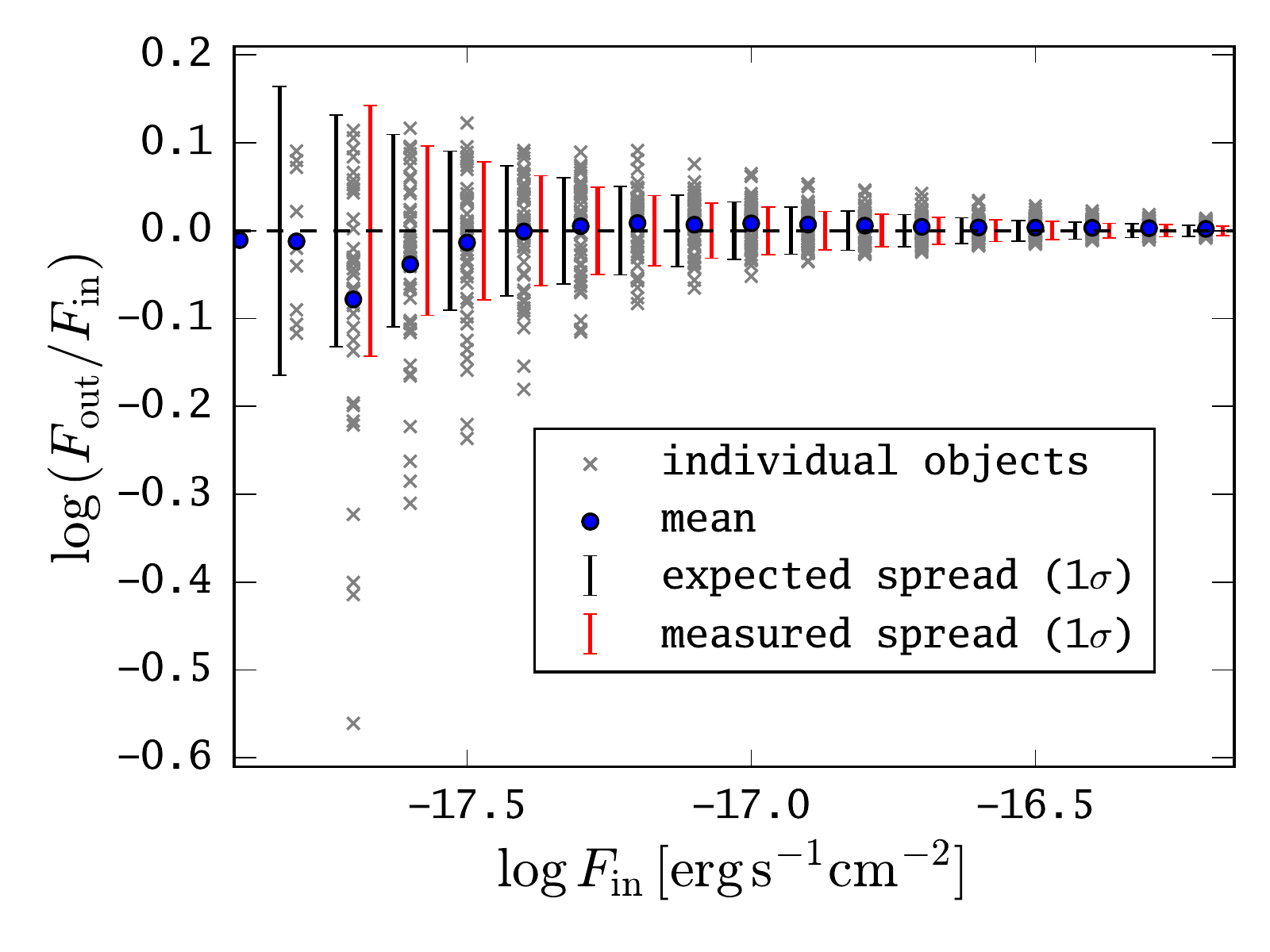}}\vspace{-1em}
  \caption{Validation of the emission line flux integration routine
    utilising implanted emission lines in the MUSE HDF-S datacube
    described in Sect.~\ref{sec:creat-test-datac}.  The input flux of the implanted emission lines is shown on the abscissa, while on
    the ordinate we show the logarithmic difference between input flux
    and measured flux by \texttt{LSDCat}:
    $\log F_\mathrm{out}\,[\mathrm{erg\,s^{-1}cm^{-2}}] - \log
    F_\mathrm{in}\,[\mathrm{erg\,s^{-1}cm^{-2}}] = \log
    (F_\mathrm{out} / F_\mathrm{in})$. Measured fluxes are obtained in
    apertures of $3\cdot R_\mathrm{Kron}$. Grey crosses show the
    obtained difference for each individual object, while the blue
    circle shows the mean difference of all measured fluxes at a given
    input flux. The red bars indicate the spread (measured standard
    deviation) over all measured fluxes at a given input flux, while
    the black bars indicate the expected spread (predicted standard
    deviation) according to the average uncertainty of the flux
    measurement at a given input flux.  For clarity, the red and black
    bars have been offset slightly to the positive and negative,
    respectively, from the actual input flux.}
  \label{fig:fluxtest}
\end{figure}

Given the known input emission line fluxes of the implanted fake
emission line sources $F_\mathrm{in}$ we are equipped to validate the
flux integration routine implemented in \texttt{LSDCat}.  As detailed
in Sect.~\ref{sec:integr-line-flux}, \texttt{LSDCat} integrates fluxes
in circular apertures of radii $k \cdot R_\mathrm{Kron}$
(Eqs.~\ref{lsd_eq:20} and \ref{lsd_eq:21}).  It has been shown that
apertures with $k=3$ are expected to contain $>99\%$ of the flux for
Gaussian profiles \citep{Graham2005}.  Therefore we compare in this
experiment $F_\mathrm{in}$ to the \texttt{LSDCat} measured flux in three
Kron-radii: $F_\mathrm{out} = F (3\cdot R_\mathrm{Kron})$.  In absence
of noise we thus expect that for every source
$F_\mathrm{out} = F_\mathrm{in}$.

The result of the above comparison from our source insertion and recovery
experiment is visualised in Fig.~\ref{fig:fluxtest} where we plot, as
a function of input flux, the difference between input and output flux
for each individual emission line source detected by \texttt{LSDCat}.
In addition to the individual differences we also show, again as a
function of input flux, the mean and the standard deviation of the
difference over all recovered sources (blue circles and red bars in
Fig.~\ref{fig:fluxtest}).  We can compare the latter with the expected
spread in flux measurements.  This expected spread is simply given by
the average uncertainty on the flux measurement as tabulated by
\texttt{LSDCat} for each input flux bin.  As noted in
Sect.~\ref{sec:integr-line-flux}, the flux measurement uncertainties
follow from direct propagation of the voxel variances through
Eq.~(\ref{lsd_eq:21}).

As can be seen from Fig.~\ref{fig:cog_vs_f3kron}, there is no bias in
the recovered flux levels above input fluxes of
$\log F_\mathrm{in} \,[\mathrm{erg\,s^{-1}cm^{-2}}] = -17.5$.  Even
for lower flux levels, the systematic errors are small, with (on
average) 90\% of the total flux being recovered.  Still, at the lowest
flux levels, where the detection completeness is also below 100\%, a
handful of implanted emission line sources are recovered with fluxes
that are only $\sim$50\%
($\log (F_\mathrm{out} / F_\mathrm{in}) \lesssim -0.3$) of the input
flux.  We checked that the larger deviations for some of the faintest
simulated sources are all due to imperfections in the HDFS datacube.
This experiment thus demonstrates at the same time the robustness of
the flux measurement procedure in LSDCat, but also the need to use
real data in quantifying the true performance of a certain measurement
approach.

\subsection{Comparison to manual flux integration on spatially
  extended objects}
\label{sec:comp-manu-flux}

\begin{figure}
  \centering
  \resizebox{\hsize}{!}{\includegraphics{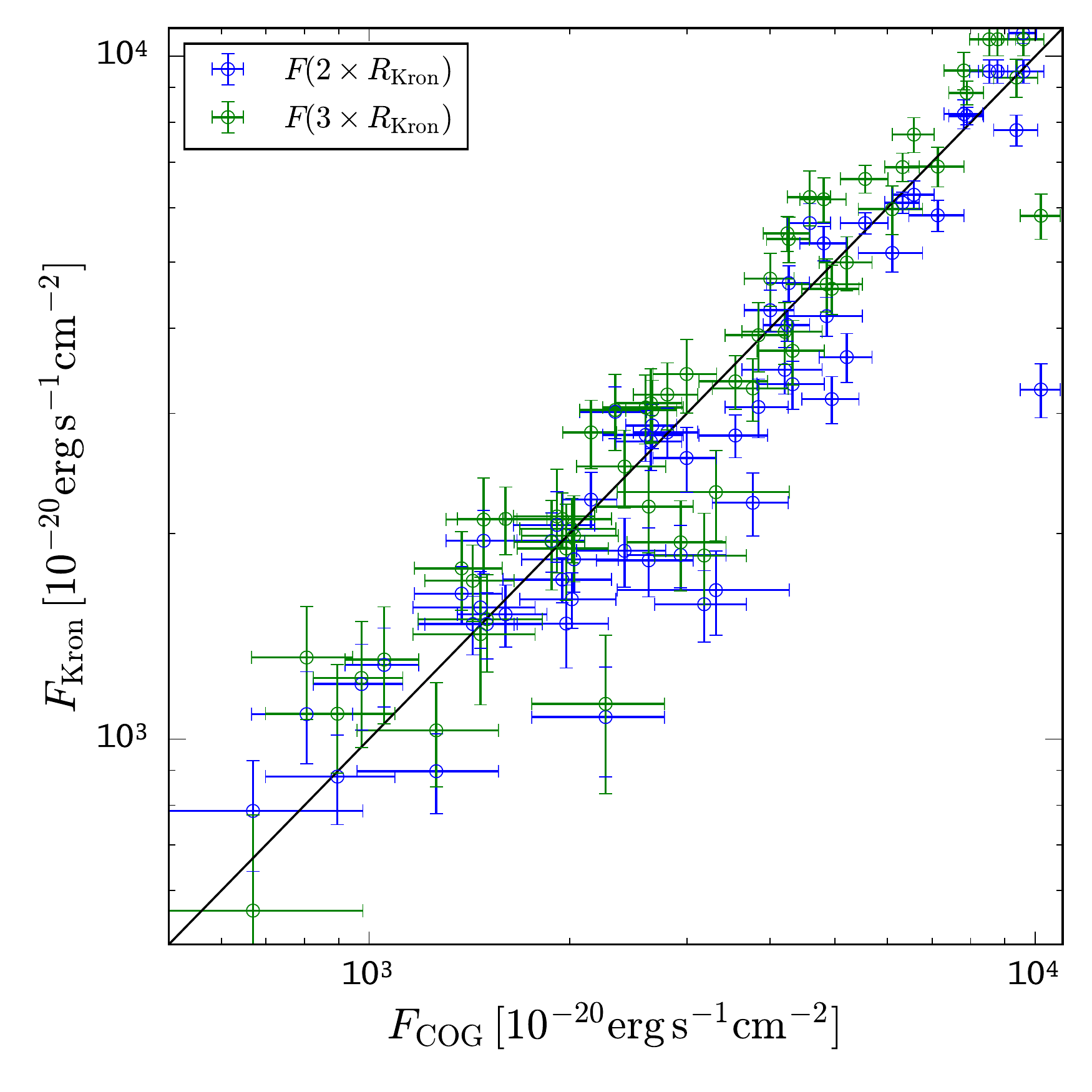}}
  \caption{Comparison of automatic flux measurements by
    \texttt{LSDCat} ($F_\mathrm{Kron}$) with fluxes measured manually
    by curve-of-growth ($F_\mathrm{COG}$) integration, for a sample of
    60 Lyman $\alpha$ emission lines in MUSE datacubes.  The automatic
    measurements were extracted according to Eq.~(\ref{lsd_eq:20}) and
    Eq.~(\ref{lsd_eq:21}), using spatial apertures of
    2$R_\mathrm{Kron}$ (blue symbols) and 3$R_\mathrm{Kron}$ (green
    symbols) and adopting $\mathit{S/N}_\mathrm{ana}=3.5$. 
      The manual method used to measure the fluxes is explained in
      Sect.~\ref{sec:comp-manu-flux}. The black diagonal line
      indicates the 1:1 relation.  The gross deviation for the
      brightest object in the sample is caused by a strongly
      double-peaked Lyman $\alpha$ line profile with significant peak
      separation. Here the narrow-band window automatically determined
      by \texttt{LSDCat} treated the stronger peak as a single
      emission line, while the trained eye was able to adjust the
      window accordingly to include both peaks.} 
  \label{fig:cog_vs_f3kron}
\end{figure}

To further demonstrate the robustness of the fluxes obtained with
\texttt{LSDCat},  we compare the flux
measurements from \texttt{LSDCat} with manually measured fluxes (Fig. 5).  The
sample on which we performed the test consists of 60 Lyman $\alpha$
emitting galaxies that were found by us in MUSE datacubes (Herenz et
al., in prep.). As established recently by \citet{Wisotzki2015}, such
galaxies show regularly extended but low-surface-brightness Lyman
$\alpha$ haloes, thus constituting good test cases for the flux
measurement in non-trivially shaped extended objects.  We obtained the
manual flux measurements by using a curve-of-growth approach on
pseudo-narrowband image created from the datacube.  The width and
central wavelength of those images were determined by eye in order to
ensure that the complete emission line signal is encompassed within
the band-pass.  Therefore we utilised 1D spectra extracted in a
circular aperture of three pixel radius centred on
$x_{\tilde{F}}^\mathrm{com}$, $y_{\tilde{F}}^\mathrm{com}$.  On these
images we then constructed the growth curve by integrating the fluxes
in concentric circular apertures with consecutively increasing radii.
Finally, we visually inspected these curves to pin down the radius at
which the curve saturates, and the total flux within this circular
aperture was adopted as the final flux measurement.  

The \texttt{LSDCat} measurements were obtained in apertures of
$R=2R_\mathrm{Kron}$ and $R=3R_\mathrm{Kron}$ and applying
Eq.~(\ref{lsd_eq:21}). As can be seen in Fig.~\ref{fig:cog_vs_f3kron},
the different measurement approaches agree very well globally. The
\texttt{LSDCat} $R=2R_\mathrm{Kron}$ fluxes are systematically
somewhat below the growth curve measurements, indicating that the
$2R_\mathrm{Kron}$ apertures still lose a small but significant
fraction of the flux (-2\,\% or -6\,\% flux lost compared to the
manual fluxes in the median or average of the sample, respectively);
this is no longer the case for the $R=3R_\mathrm{Kron}$ apertures
(+10\,\% or +8\,\% flux gained in the median or average,
respectively).  We conclude that \texttt{LSDCat} delivers reliable and
robust flux measurements for emission lines with non-pathological
spatial and spectral shapes.

\section{Guidelines for the usage of \texttt{LSDCat}}
\label{sec:gudel-optim-param}

We now provide some guidelines for using \texttt{LSDCat} on
wide-field IFS datacubes.  These are based on our experience with
applying the code on MUSE datacubes, searching for
faint emission lines from high-redshift galaxies (first results presented
in \citealt{Bacon2015} and \citealt{Bina2016}; more will be reported
in Herenz et al., in prep).  
We intend these guidelines to be instructive for the potential
\texttt{LSDCat} user, but they should not be understood as recipes.

\subsection{Dealing with bright continuum sources in the datacube}
\label{sec:dealing-with-bright}

Even in relatively empty regions in the sky, any blank-field exposure 
will contain objects that produce a detectable continuum signal within 
the datacube, and that correspondingly appear in a 
white-light image resulting from averaging over all layers in the 
datacube (e.g. Fig.~3 in \citealt{Bacon2015}).  
To detect such sources, conventional 2D source detection
algorithms are clearly sufficient. Since in the
detection process of \texttt{LSDCat} we implicitly assume any significant 
signal to be due to emission lines, it is advisable to either mask out
or, better, subtract any significant continuum signal from the datacube
before running \texttt{LSDCat}. This step is not strictly needed, and 
the presence of continuum sources in the datacube does not render 
the detection algorithm unusable as such. But the significance of
a line detection would clearly change if the line sat on top of a continuum
signal, while a very bright continuum-only object would even turn out a 
band of spurious detections.

To remove the continuum, we found it useful to create and subtract
a ``continuum-only'' cube by median filtering the original flux datacube 
in spectral direction only. The width of the median filter should be much 
broader than the expected widths of the emission lines, and it should 
be narrow enough to approximately trace a slowly varying
continuum.  In our experience, filter radii of $\sim150$\,\AA{}--$200$\,\AA{} 
serve these goals very well.  The median filter-subtracted datacube can 
then be used as input $\bm{F}$ for \texttt{LSDCat}.

\subsection{Width of the spatial filter template}
\label{sec:optim-choice-seeing}

\begin{figure}
  \centering
  \resizebox{\hsize}{!}{\includegraphics{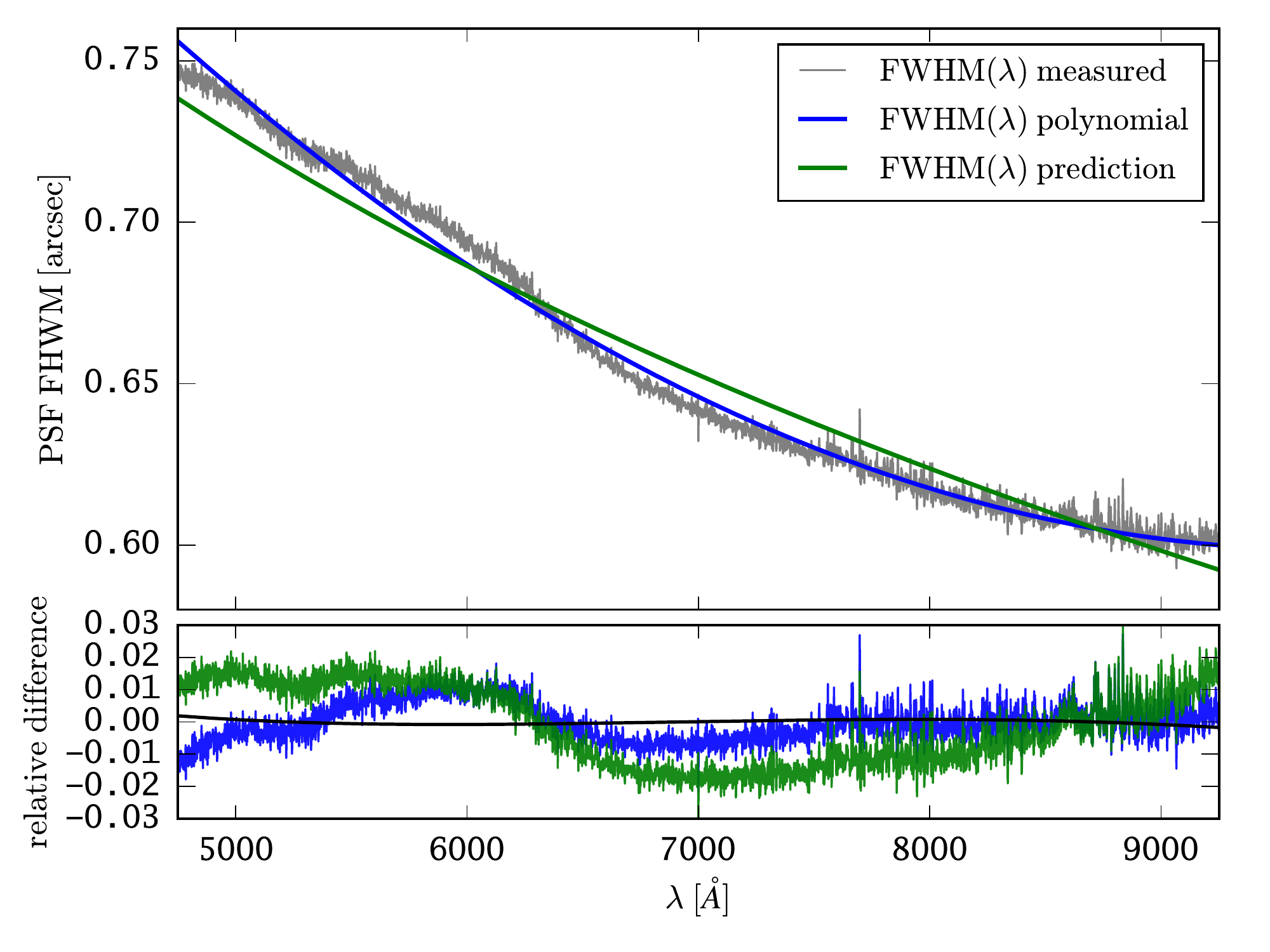}}
  \caption{Wavelength dependence of the PSF FWHM in the MUSE HDFS
    datacube \citep{Bacon2015}.  The grey curve shows the FWHM from a
    \citeauthor{Moffat1969} fit to the brightest star in the field,
    the blue curve shows a quadratic polynomial fit to the grey curve,
    and the green curve shows the analytic prediction for
    $\mathrm{FWHM}(\lambda)$ from \cite{Tokovinin2002} (see
    Sect.~\ref{sec:optim-choice-seeing} for details).  In the bottom
    panel we show the relative difference between polynomial and
    analytic prediction from the measured PSF FWHM.  The black curve
    in the bottom panel shows the relative difference between the
    analytic prediction and a polynomial fit to this prediction.}
  \label{fig:fwhm_lambda}
\end{figure}

The matched filtering approach produces maximum significance if the 
shapes of the true signal and of the template exactly agree; any
template mismatch leads to a reduced $\mathit{S/N}_\mathrm{peak}$ 
value. Fortunately, this dependence of $S/N$ on the template shape parameters
is relatively weak around the optimum. If both signal and template are of Gaussian
shape and the template has an incorrect width of 
$\mathrm{FWHM}_\mathrm{templ} = \kappa \times\mathrm{FWHM}_\mathrm{true}$,
it can be shown that $\mathit{S/N}_\mathrm{peak}$ decreases only as
\begin{equation}   
   \label{eq:zackay}
   \mathit{S/N}_\mathrm{peak} \propto 2\kappa / (\kappa^2 +1 )
\;\text{,}\end{equation}
\citep[][]{Zackay2015}.  Hence even a difference of 20\% between the adopted and the correct FWHM will result in a reduction of $S/N$ by only $\sim$2\%, entirely negligible for our purposes; this number is supported by our own numerical experiments.

When searching for compact emission line objects, a single spatial template
modelling the light distribution of a point source (i.e.\ the PSF) will therefore 
be sufficient for many applications. Even neglecting the wavelength dependence
of the seeing (i.e.\ setting the polynomial coefficient $p_0$ in Eq.~(\ref{eq:2}) 
to the mean Gaussian seeing FWHM, and all other PSF parameters to zero)
will result in only a very modest reduction of sensitivity at the lowest and highest 
wavelengths.

To go one step further in accuracy, one has to account for the seeing as a 
function of wavelength.
In the framework of the standard (Kolmogorov) turbulence model of the
atmosphere, the seeing is expected to decrease with increasing
wavelength as $\mathrm{FWHM} \propto \lambda^{-1/5}$, where the
constant of proportionality also depends on airmass \citep[e.g.][]{Hickson2014}. 
In reality, other effects also play a role, such as guiding errors, or the blurring
induced by co-adding multiple dithered exposures of the same field.

When the observed field contains a point source of sufficient brightness, 
a direct measurement of $\mathrm{FWHM}(\lambda)$ 
is possible from the datacube.  
As an example, we show in Fig.~\ref{fig:fwhm_lambda} the derived
$\mathrm{FWHM}(\lambda)$ relation from a \citeauthor{Moffat1969} fit
to the brightest star in the MUSE HDFS datacube 
\citep[same data as Fig. 2 in][]{Bacon2015}. Here we overplot the 
polynomial fit according to Eq.~(\ref{eq:2}) and the relative difference between
this fit and the $\mathrm{FWHM}$ values derived from the star,
$(\mathrm{FWHM}_\mathrm{star} - \mathrm{FWHM}_\mathrm{fit}) /
\mathrm{FWHM}_\mathrm{star}$.  This difference is less
than 3\%, thus totally negligible in our matched-filter application.  For
datacubes containing only faint stars, it
may be advisable to bin over several spectral layers before 
modelling the PSF.

If no point source is present within the datacube, the relations 
by \cite{Tokovinin2002} can be used to get an idea of the dependence
of FWHM on $\lambda$. They predict $\mathrm{FWHM}(\lambda)$
given a differential image motion monitor (DIMM) seeing FWHM
measurement, the airmass of the observation, and an additional
parameter $\mathfrak{L}_0$ (called the wavefront outer scale length)
that quantifies the maximum size of wavefront perturbations 
by the atmosphere \citep[e.g.][]{Martin1998}.
In Fig.~\ref{fig:fwhm_lambda} we also compare the 
$\mathrm{FWHM}(\lambda)$ relation derived from this model to 
the actual measurement of the brightest star in the HDFS.  
For the plot, we adopted a DIMM seeing of 0.75\arcsec{} at 5000\AA{} 
and an airmass of 1.41, both averages over all individual MUSE HDFS
observations.  We set $\mathfrak{L}_0$ to 22\,m, which is the
median of this for the Paranal observatory \citep{Conan2000}.  
As can be seen, this model provides a very good
description of the measured $\mathrm{FWHM}(\lambda)$ relation.
Moreover, from this figure it is also clear that a second-order polynomial 
is a nearly perfect representation of the \citeauthor{Tokovinin2002} 
$\mathrm{FWHM}(\lambda)$ prediction.

\subsection{Width of the spectral filter template}
\label{sec:optim-choice-v_mathr}

\begin{figure}
  \centering
  \resizebox{\hsize}{!}{\includegraphics{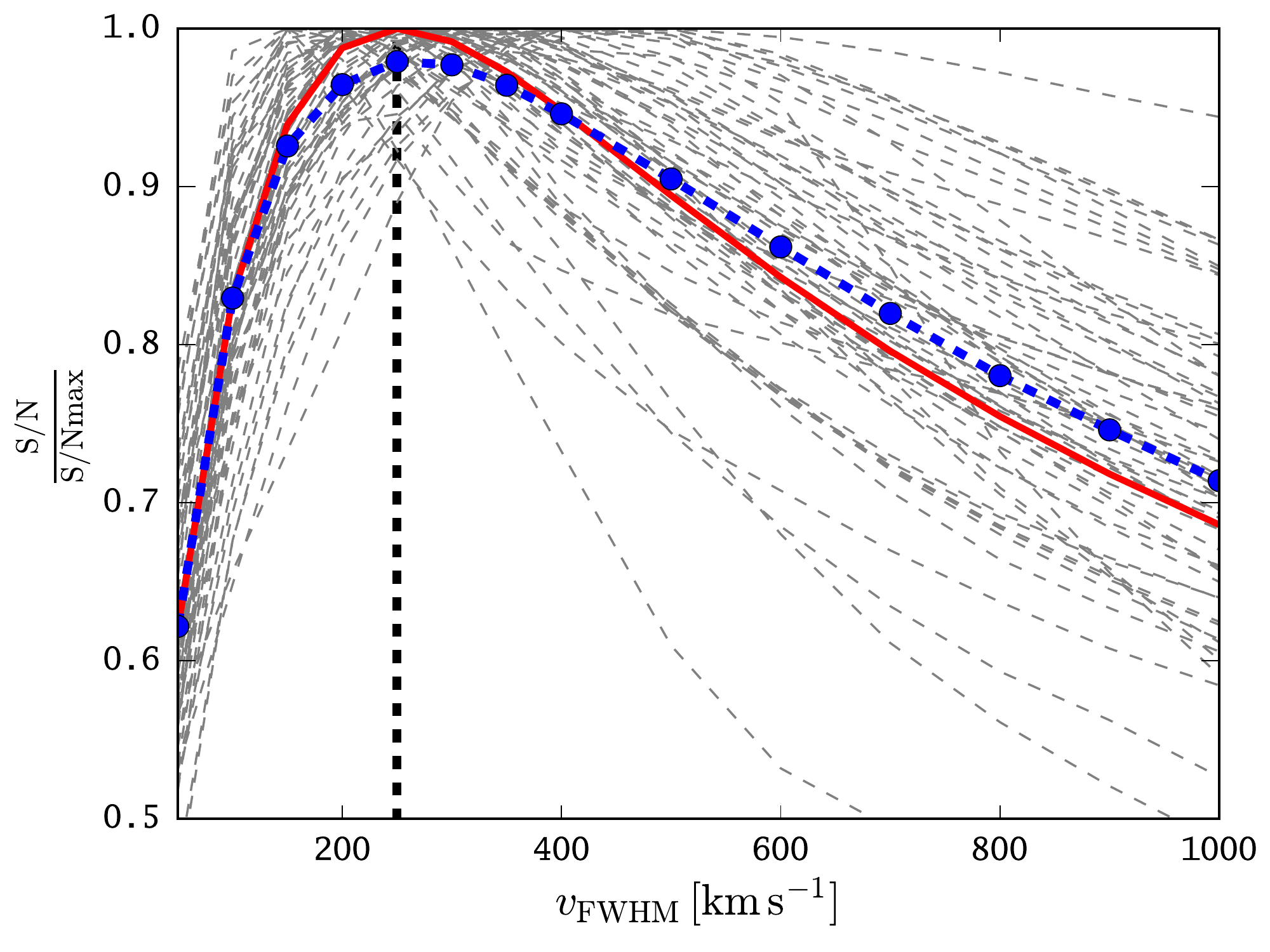}} \vspace{-1.5em}
  \caption[]{Ratio $\xi = (\mathit{S/N})/(\mathit{S/N}_\mathrm{max})$ for
    all Lyman $\alpha$-emitting sources in the Hubble Deep
    Field South MUSE datacube from
    \cite{Bacon2015}. Here, $\mathit{S/N}_\mathrm{max}$ is the maximum
    $S/N_\mathrm{peak}$ value of a source, over all considered filter
    widths. The grey lines denote the $\xi$ for the individual emitters, while the blue
    points and connecting dashed curve show the average 
    relation.  For a filter width of
    $\mathrm{FWHM}_v = 250$\,km\,s$^{-1}$ (vertical dashed line),
    almost all Lyman $\alpha$ line emitters have $\xi > 90$\%.  The red
    curve shows the theoretically expected ratio
    $\xi \propto \sqrt{2\kappa / \kappa^2+1}$ 
    for an assumed Gaussian emission line with
    $\mathrm{FWHM}_{v,\mathrm{true}}=250$\,km\,s$^{-1}$ filtered with
    an incorrect template of width
    $\mathrm{FWHM}_v = \kappa \times\mathrm{FWHM}_{v,\mathrm{true}}$.  
    }
  \label{fig:vfwhm_filt}
\end{figure}

As explained above, \texttt{LSDCat} assumes the spectral (Gaussian)
template to have a fixed width in velocity space. We now briefly
demonstrate the effect of template mismatch in the spectral
domain. Similarly to the 2D case considered in the previous subsection
(Eq.~\ref{eq:zackay}), it can be shown that
$\mathit{S/N}_\mathrm{peak}$ decreases as
\begin{equation}
  \label{eq:lsd_own}
  \mathit{S/N}_\mathrm{peak} \propto \sqrt{2\kappa/(\kappa^2+1)}\;\text{,}
\end{equation} 
where $\kappa$ is the ratio between adopted and true template width. Even when the filter width is half or twice that of the
actual object, the maximum reachable detection significance
reduces by only $\sim 10\%$.  For the same reason, the achievable
$S/N$ is very robust against moderate shape mismatches between real
emission line profiles and the Gaussian template profile.

A good choice of the spectral filter width $\mathrm{FWHM}_v$ can be
motivated by analysing the distribution of expected emission line
widths (taking instrumental line broadening into account).  
If the distribution of observed line widths is relatively narrow 
it will be sufficient to adopt a single template with width
close to the midpoint of the expected distribution. If however the expected distribution
is very broad, for example, when searching both for star-forming galaxies and for AGN, 
it may be useful to generate two filtered datacubes with
significantly different $\mathrm{FWHM}$ values (the ratio should be at least a factor 3).
This implies two \texttt{LSDCat} runs creating two catalogues that later have to be merged. 

To demonstrate the impact of varying the template width
$\mathrm{FWHM}_v$ on the detectability of a particular class of
emission-line objects we show in Fig.~\ref{fig:vfwhm_filt} the
dependence of the ratio
$\xi \equiv (\mathit{S/N})/(\mathit{S/N}_\mathrm{max})$ for all Lyman
$\alpha$ emitters in the MUSE Hubble Deep Field South datacube
\citep{Bacon2015}.  Here, $\mathit{S/N}_\mathrm{max}$ denotes the
maximum detection significance for a line, comparing all considered
line widths. The plot shows that $\xi$ varies quite slowly with
template width, confirming the theoretically expected behaviour (shown
by the red curve). A good choice, at least for this object class,
appears to be a filter width of $\mathrm{FWHM}=250$\,km\,s$^{-1}$, for
which basically all the Lyman $\alpha$ emitters in the sample are
detected with at least 90\% of their maximum possible detection
significance. The same template will capture even completely
unresolved emission (i.e.\ with just the MUSE instrumental line width
FWHM of $\sim$100\,km\,s$^{-1}$ at $\lambda = 7000$\,\AA) at more than
80\% of the value for a perfectly matching template.

\subsection{Empirical noise calibration}
\label{sec:effective-variances}

Source detection is essentially a decision process based on a
test statistic to either reject or accept features in the data as
genuine astronomical source signals \citep[e.g.][]{Schwartz1975,Wall1979,Hong2014,Zackay2015,Vio2016}.  This
decision is usually based on a comparison with the noise statistics of
the dataset under scrutiny.  Consequently a good knowledge of
the noise properties is required for deciding on meaningful
thresholds, for example, in terms of false-alarm probability.  In
\texttt{LSDCat} we {assume} that $\bm{\sigma}^2$ contains a good
estimate of the variances. However, in reality this is often not so easy
to obtain. In particular, resampling processes carried during the data
reduction usually neglect the covariance terms; even if known, it 
would be computationally prohibitive to formally include them in a
dataset of $4\times 10^8$ voxels. In consequence, any direct 
noise property derived from $\bm{\sigma}^2$ will underestimate 
the true noise, possibly by a very substantial factor, and the
resulting detection significances will be biased towards overly 
high values that lose their probabilistic connotation. 

A possibly remedy is self-calibration of the noise from the flux datacube.
There are several ways to accomplish this, and here we simply provide a few
insights from our own experience with MUSE cubes. It must be
realised that the pixel-to-pixel noise in any given spectral layer will be
as much affected by the resampling as the propagated variances, and 
therefore biased low in the same way. This is not so, however, for the variance
of a typical {aperture}, for which the effects of resampling are much
lower (basically acting only on the pixels at the circumference of the
aperture). One can therefore estimate {effective variances} by evaluating
the standard deviation between several identical apertures placed in 
different blank sky locations, separately for each spectral layer. 
This approach  also implicitly accounts
for additional noise-like effects due to imperfect flatfielding or 
sky subtraction. 

Indeed, sky subtraction residuals can still be an issue in MUSE data, especially for 
the OH forest in the red part of the spectral range. If a dataset 
should be heavily affected and suffer from too many spurious detections that
are just sky residuals (and even the new ZAP software by \citealt{Soto2016}
not providing sufficient improvement), it is still possible to increase the effective
variances for the affected spectral layers to a level where only extremely bright
sources at these wavelengths are detected by \texttt{LSDCat}.

\subsection{Choosing the detection and analysis thresholds}
\label{sec:optim-choice-detect}

\begin{figure}
  \centering
  \resizebox{\hsize}{!}{\includegraphics{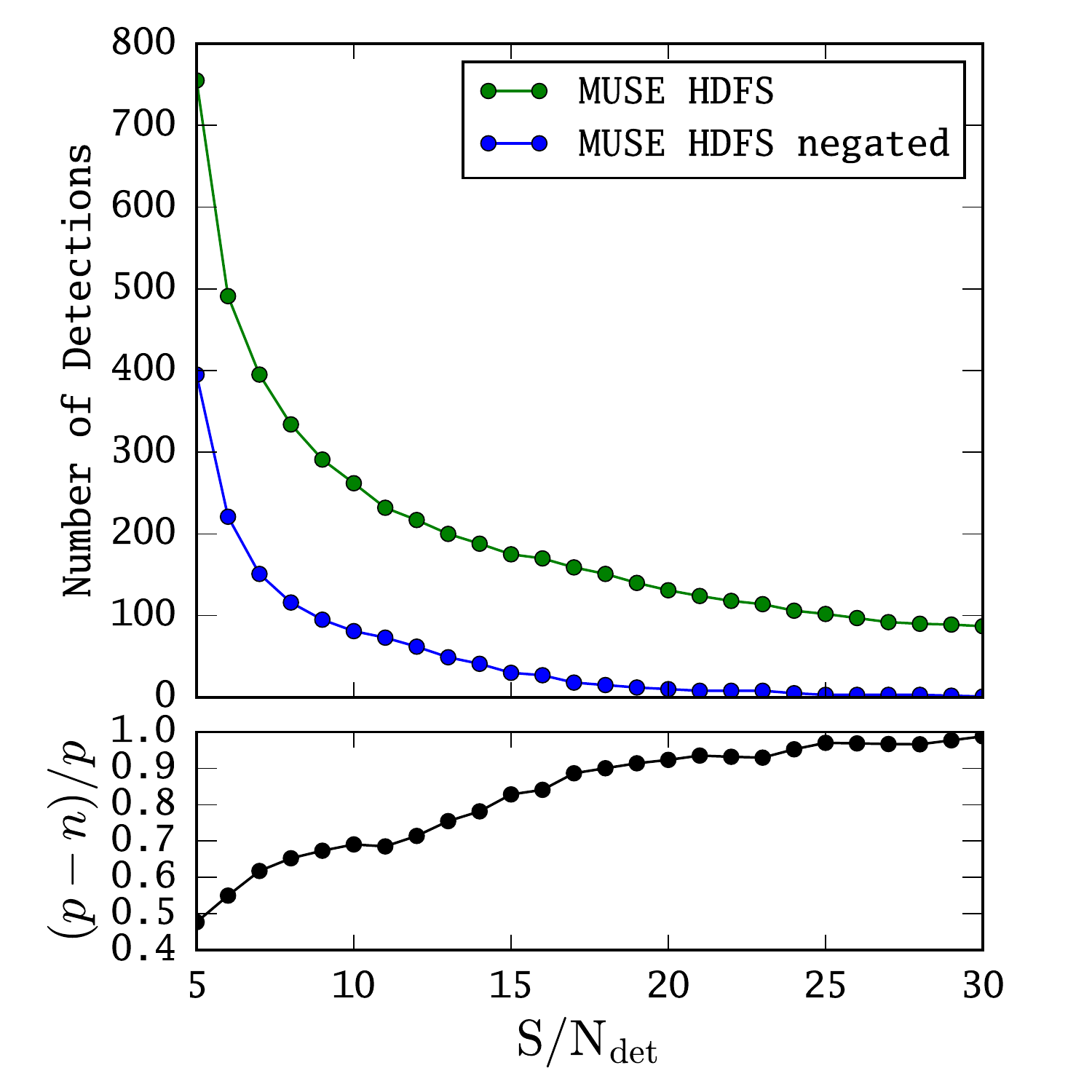}}
  \vspace{-1.5em}
  \caption{\emph{Top panel}: Number of \texttt{LSDCat} emission line
    detections at different $\mathit{S/N}_\mathrm{det}$ thresholds
    in the MUSE HDFS datacube (green symbols) and in a negated
    version of it (blue symbols).  \emph{Bottom panel}: Expected rate of
    genuine detections approximated by the ratio $R=(p-n)/p$, where $p$
    is the number of emission line detections in the MUSE HDFS
    datacube and $n$ is this number in the negated dataset.}
  \label{fig:hist}
\end{figure}

A crucial quantity in each detection process is the incidence rate of
false positives. If the noise properties are accurately known, the
false-alarm probability can be directly calculated from the detection
threshold $\mathit{S/N}_\mathrm{det}$. Given the $\gtrsim 10^8$ voxels
in a MUSE-like wide-field datacube it is obviously necessary to have a
very low false-alarm probability. After matched filtering with a
typical template the number of independent voxels gets reduced by a
factor of $\sim10^2$. (Here we consider as a typical template
$v_\mathrm{FWHM} = 300$\,km\,s$^{-1}$ and $p_0=0.8\arcsec{}$ (the
default values in \texttt{LSDCat}). Given the spatial sampling of
0.2\arcsec{} per spatial pixel and spectral sampling of 1.25\AA{} per
datacube layer in MUSE, this corresponds at the central wavelength
range of MUSE ($\sim$6980\AA{}) to a filter FWHM of $\sim$10 voxels.)
Even assuming perfect Gaussian noise, a threshold of
$\mathit{S/N}_\mathrm{det} = 5$ would then already result in $\sim10$
spurious detections per cube.  However, the wings of the noise
distribution are never perfectly Gaussian, to which also flatfielding
and sky subtraction residuals have to be added; any such deviations
from Gaussianity will directly inflate the false detection rate.

Instead of relying on theoretically expected false alarm
probabilities, we recommend self-calibrating the rate of spurious
detections using \texttt{LSDCat} on the negated cube $-\bm{F}$.
While 
obviously no real emission line object can be detected in such a
dataset, it can probably be assumed that the effective noise
properties are approximately the same in original and negated
data. Running \texttt{LSDCat} on both versions then immediately allows
us to estimate the ratio of real to spurious detections and adjust the
final value of $\mathit{S/N}_\mathrm{det}$ accordingly. A possible
criterion could then be the \emph{point of diminishing returns}, where
lowering the detection threshold would produce a large increase in
spurious detections with only a small compensatory increase of genuine
emission lines.

In Fig.~\ref{fig:hist} we exemplarily present the result of such an
analysis for the MUSE HDFS dataset.  This datacube was processed
according to the guidelines given in the previous subsections.  In the
figure, we show both the number of detections in the original- and in
the negated datacube at different thresholds.  Moreover, in the bottom
panel of Fig.~\ref{fig:hist} we show the ratio $R=(p-n)/p$, where $p$
and $n$ designate the number of detections in the original and negated
datacube, respectively.  Assuming a symmetric noise distribution
around zero and no systematic negative holes in the data, the rate of
genuine detections within a catalogue at a given threshold can be
approximated by $R$ \citep[e.g.][]{Serra2012}.  However, we find that
the sky-subtraction residuals in the MUSE HDFS datacube appear to be
systematically skewed to more negative values.  In addition, since we
subtracted continuum signal utilising the median filter approach
(Sect.~\ref{sec:dealing-with-bright}), absorption lines in some
continuum bright objects created holes that mimic emission line
signals in the negated datacube.  Indeed, most of the detections in
the negated cube at $\mathit{S/N}_\mathrm{det}\gtrsim10$ can be
associated either with sky subtraction residuals or holes from
over-subtracted absorption lines.  Taking this into account, we notice
in Fig.~\ref{fig:hist} a second strong increase of detections in the
negated dataset that is not compensated by positive detections at
$\mathit{S/N}_\mathrm{det}\lesssim 8$.  Hence, for this particular
dataset a threshold below $\mathit{S/N}_\mathrm{det}\approx 8$ would
not be advisable.

The choice of the analysis threshold $\mathit{S/N}_\mathrm{ana}$ used
in the measurement routine (Sect.~\ref{sec:meas}) again depends on the
noise characteristics. Here it is useful to visually inspect the
$\bm{S/N}$-cubes.  Considering for example the faint emission
line source presented in Fig.~\ref{fig:detsnexample}, by comparing the
left panel (where no source is present) to the right panel that
includes the source, we find that non-source voxels rarely obtain
$S/N$ values high than 3.  Hence, in this dataset, voxels around a
real detection above a threshold value of 3.5 are likely linked to the
actual source and should be included in the measurement process.

\section{Conclusion and outlook}
\label{sec:conclusion}

Here we present \texttt{LSDCat}, a conceptually simple but
robust and efficient detection package for emission line objects in
wide-field IFS datacubes.  The detection utilises a 3D
matched-filtering approach to detect individual emission lines and
sorts them into discrete objects.
Furthermore, the software measures fluxes and the spatial 
extents of detected lines. \texttt{LSDCat} is implemented in Python, 
with a focus on fast processing of the large data volumes generated
by instruments such as MUSE.  In this paper we also provide some 
instructive guidelines for the prospective usage of \texttt{LSDCat}.

\texttt{LSDCat} is open-source.  Following the example of AstroPy
\citep{AstropyCollaboration2013}, we release it to the community under
a 3-clause BSD style
license\footnote{\url{https://www.w3.org/Consortium/Legal/2008/03-bsd-license.html}}.
This license permits usage and modification of the code as long as
notice on the copyright holders (E.C. Herenz \& L. Wisotzki) is given.
A link to download the software is provided on the \emph{MUSE Science
  Web Service} at \url{http://muse-vlt.eu/science/tools/}, and it is
also available via the Astrophysics Source Code Library at
\url{http://ascl.net/1612.002} \citep{Herenz2016a}.

\texttt{LSDCat} is documented in two ways. First, each of the routines
described in Sect.~\ref{sec:method-description} is equipped with an online
help.  Secondly we provide an extensive README file describing all
the routines and options in detail.  Moreover, that README contains
examples and scripts that help use \texttt{LSDCat} efficiently.

\texttt{LSDCat} is actively maintained as it is currently used by the
MUSE consortium to search for high-$z$ faint emission line galaxies
(e.g., \citealt{Bacon2015}, \citealt{Bina2016}, Herenz et al., in
prep., Urrutia et al., in prep.).  Development takes place within a
git repository\footnote{\url{http://git-scm.com}}. Technically
inclined members of the community are invited to contribute to the
code.  We also offer a bug tracker that allows users to report
problems with the software.

While \texttt{LSDCat} is fully operational, we see a number of aspects
where there is room for future improvement.  For example,
\texttt{LSDCat}  currently does not perform a deblending of over-merged
detections, nor does it automatically merge detections belonging to a
single source unless their initial positions are within a predefined
radius. Indeed, in our search for faint line emitters in MUSE
datacubes we encountered a few cases of very extended line-emitting
galaxies that fragmented into several `sources'.  We aim at
addressing this problem in a future version.  Currently these sources
have to be merged or deblended manually in the resulting output
catalogue.  Another improvement planned for a future release is an
automatic object classification for objects where multiple lines are
detected.  To this aim, the combination of spectral lines found at the
same position on the sky must match a known combination of redshifted
galaxy emission line peaks \citep{Garilli2010}.  Finally, while in
principle the software could be used for any sort of astronomical
datacubes (e.g. coming from radio observations, or other IFS
instruments), we have so far focused our efforts in development and testing
on MUSE datacubes.  However, the code is independent of instrument
specifications and requires only valid FITS files following the
conventions stated in Sect.~\ref{sec:input}.  Still, despite all these
possible enhancements \texttt{LSDCat} is already a complete software
package, and we hope that it will be of value to the community.

\begin{acknowledgements}
  We thank Maria Werhahn for valuable help with the parameter study
  shown in Fig.~\ref{fig:vfwhm_filt} and Joseph Caruana for providing
  us with curve-of-growth flux integrations of 60 Ly$\alpha$ emitters
  shown in Fig.~\ref{fig:cog_vs_f3kron}.  We thank Rikke Saust for
  preparing the test data.  We also thank the MUSE consortium lead by
  Roland Bacon for constructive feedback during the \texttt{LSDCat}
  development.  E.C.H. dedicates this paper to Anna's cute fat cat
  \emph{Cosmos}.
\end{acknowledgements}

\bibliographystyle{aa}
\bibliography{lsdcatpaper.bib}

\appendix

\section{Usage example}
\label{sec:usage-example}

We provide a short example to demonstrate how the user interacts with
the \texttt{LSDCat} routines to obtain an emission line catalogue from
an IFS datacube (see also Fig.~\ref{fig:lsdcat_flow}).  For this
example we use the MUSE HDFS datacube \citep{Bacon2015} in version
1.34. This datacube can be downloaded from the MUSE consortium data
release web page: \url{http://muse-vlt.eu/science/data-releases/}.
For brevity, we refer to the MUSE HDFS datacube FITS file as
\texttt{datacube.fits}.  Moreover, in all following commands, the
\texttt{-o} option specifies the output filename of a particular
routine.

As detailed in Sect.~\ref{sec:dealing-with-bright} any detectable
source continua should be removed from a datacube on which
\texttt{LSDCat} is used.  There we outlined, that it is often
sufficient to subtract these continua by subtracting an
in-spectral-direction median filtered version of the datacube from the
original datacube.  We package with \texttt{LSDCat} an additional
routine \texttt{median-filter-cube.py} that performs this task via the
following command:
\begin{verbatim}
median-filter-cube.py datacube.fits \
 -o mf_datacube.fits
\end{verbatim}
(Run with its default settings the full width of the running median is
180\AA{}.)  The output file \texttt{mf\_datacube.fits} contains
suitable input datacubes $\bm{F}$ and $\bm{\sigma}^2$ for the
matched-filtering procedure (Sect.~\ref{sec:3d-matched-filtering}).

The spatial convolution and corresponding error propagation can now be
achieved by running the following command:
\begin{verbatim}
lsd_cc_spatial.py  --gaussian  \
 -p0=0.65 -p1=-4.5e-5 --lambda0=7050 \
 -i  mf_datacube.fits  -m mask.fits \
 -o spac_datacube.fits
\end{verbatim}
Here we specified with \texttt{-p0}, \texttt{-p1} and
\texttt{{-}{-}lambda0} the coefficients and zero-point of the linear
function
$p(\lambda [\mathrm{\AA}]) = 0.65 - 4.5\times10^{-5}(\lambda
[\mathrm{\AA}] - 7050\mathrm{\AA})$ which is an apt representation of
the PSF FWHM wavelength dependency in this field.  The switch
\texttt{{-}{-}gaussian} models the PSF as a circular Gaussian
(Eq.~\ref{lsd_eq:8}).  Here we also utilise a mask \texttt{mask.fits}
that masks out imperfections in the HDFS datacube near the FoV borders
and the brightest star in the field, as these features cause numerous
unwanted detections in the datacube\footnote{In order to follow the
  example we provide a suitable mask for the HDFS datacube version
  1.34 in the \texttt{examples} folder of the \texttt{LSDCat}
  repository.}.  In the \texttt{LSDCat} documentation, we detail how
such a mask can be created.  The output file is named
\texttt{spac\_datacube.fits}.

Next, the spectral convolution is run on \texttt{spac\_datacube.fits} via
\begin{verbatim}
lsd_cc_spectral.py -i spac_datacube.fits \
 --FWHM=300 -o 3d_filtered_datacube.fits
\end{verbatim}
Here \texttt{{-}{-}FWHM} is used to specify a filter FWHM of
300\,km\,s$^{-1}$.  The output \texttt{3d\_filtered\_datacube.fits}
 now contains the matched filter output $\bm{\tilde{F}}$ and
$\bm{\tilde{\sigma}}^2$ from which we can compute the signal-to-noise
cube $\bm{S/N}$ (Eq.~\ref{lsd_eq:12a}) via:
\begin{verbatim}
s2n-cube.py -i 3d_filtered_datacube.fits \
 -o sncube.fits
\end{verbatim}
While optional, the above step reduces the execution time of the routines \texttt{lsd\_cat.py} and
\texttt{lsd\_cat\_measure.py}, as they normally would create the
$\bm{S/N}$ datacube on the fly.

We can now create an emission line source catalogue with
$S/N_\mathrm{det} > 10$ by typing:
\begin{verbatim}
lsd_cat.py -i sncube.fits -t 10 -c catalogue.cat
\end{verbatim}
Utilising the thresholding procedure described in
Sect.~\ref{sec:thresh-gener-interm}, the above command creates a ASCII
and FITS table \texttt{catalogue.cat} and \texttt{catalogue.fits}
for 2456 emission line candidates from 225 potential individual
objects.  Finally, the basic measurements described in
Sect.~\ref{sec:meas} are obtained for each of the emission line candidates
with the command
\begin{verbatim}
lsd_cat_measure.py -ic catalogue.fits -ta 5 \
 -f mf_datacube.fits \
 -ff 3d_filtered_datacube.fits \
 -ffsn sncube.fits
\end{verbatim}
At this stage we recommend the catalogue to be inspected with
QtClassify\footnote{\url{https://bitbucket.org/Leviosa/qtclassify}}, a
graphical user interface that helps to classify emission line candidates in IFS data, and
which is optimally suited to work with catalogues created by
\texttt{LSDCat} (Kerutt, ASCL submitted).

More complete and up-to-date documentation describing all the routines
can be found in the \texttt{LSDCat} repository.  Moreover, for all
routines, on-line usage information can be displayed by calling them
with the \texttt{-h} switch.

\end{document}